\documentclass[12pt,a4paper]{article}
\usepackage[demo]{graphicx}
\usepackage{subcaption}
\usepackage{t1enc}
\usepackage[latin2]{inputenc}
\usepackage{amsmath}
\usepackage{bbm}
\usepackage[none]{hyphenat}
\usepackage{natbib}
\usepackage{setspace}
\usepackage{tikz}
\usetikzlibrary{patterns}
\usepackage{multirow} 
\usepackage{geometry}
\usepackage{graphicx}
\usepackage{float}
\usepackage{amsthm}
\usepackage{amsfonts}
\usepackage{pgfplots}
\pgfplotsset{compat=1.12}
\usepgfplotslibrary{fillbetween}
\usepackage{pdflscape}
\usepackage{rotating}
\newtheorem{tetel}{Theorem}
\newtheorem{prop}[tetel]{Proposition} 
\newtheorem{remark}[tetel]{Remark}
\newtheorem{corol}[tetel]{Corollary}

\DeclareMathOperator{\sgn}{sgn}

\newcommand{\red}{\textcolor{black}}
\newcommand{\green}{\textcolor{black}}

\begin{document}

\sloppy

\title{Bertrand oligopoly in insurance markets with Value at Risk Constraints}

\date{}
\maketitle

\begin{abstract}
Since 2016 the operation of insurance companies in the European Union is regulated by the Solvency II directive. According to the EU directive the capital 
requirement should be calculated as a 99.5\% of Value at Risk. In this study, we examine the impact of this capital requirement constraint on equilibrium premiums and capitals. We discuss the case of the oligopoly insurance market using Bertrand's model, assuming profit maximizing insurance companies facing Value at Risk constraints. First we analyze companies' decision on premium level. The companies strategic behavior can result positive as well as negative expected profit for companies. The desired situation where
competition eliminate positive profit and lead the market to zero-profit state is rare. Later we examine ex post and ax ante
capital adjustments. Capital adjustment does not rule out market anomalies, although somehow changes them. Possibility of capital
adjustment can lead the market to a situation where all of the companies suffer loss. Allowing capital adjustment results
monopolistic premium level or market failure with positive probabilities.

\end{abstract}

\textbf{Authors:} Veronika Varga, \textit{Corvinus University of Budapest} and Kolos Csaba Ágoston, \textit{Corvinus University of Budapest}
\section{Introduction} \label{sec:intro}

After the  financial crisis of 2008 the regulation of financial institutions became a central issue. Since 2016 the operation of insurance companies in the European Union (EU) is regulated by Solvency II. This EU directive ensures the same regulation for all member states and the protection of policy holders and beneficiaries. According to Solvency II directives the solvency capital requirement of insurance companies should ensure that bankruptcy occurs not more often than once in every 200 cases \citep{solvency}. So capital is able to cover losses in 99.5\% of cases. The level of the capital requirement plays a crucial role in the operation of insurance companies. 

The main objective of this paper is to analyze the effect of the solvency capital requirement on equilibrium prices, and to determine which factors influence equilibrium prices. The natural approach is to model the sector as a Bertrand oligopoly since insurers decide on prices \citep {SonnenholznerWambach}. In the classic Bertrand model two companies are already enough to achieve the same premium and output at equilibrium which coincides with those in perfect competition. This phenomenon is called the Bertrand paradox. There are several modifications of the Bertrand model when companies can achieve positive profits and equilibrium prices are higher. Assuming premium matching guarantees in a Bertrand model leads to a continuum  of symmetric Nash equilibria \citep{Dixit}, and in several cases companies achieve positive profits. \citet {Polborn} and \citet {Wambach} have shown that the Bertrand paradox does not hold in insurance markets and positive profits can be achieved. These phenomena are attributed to the specialties of the insurance market, mainly to risk aversion of the firms and uncertain claims.  

The effect of the solvency capital requirement can also cause some anomalies in the market equilibrium. We assume that insurance companies maximize their expected profits. Profit maximization of insurers (risk neutrality) is often assumed in economic models, see for instance \citet{Schlesinger}. 

According to Solvency II directives, the solvency capital requirement should ensure that bankruptcy occurs not more often than once in every 200 cases. Alternatively, with a probability of at least 99.5 \% the insurance companies meet their obligations to policy holders and beneficiaries over the following 12 months. This capital requirement can be calculated as a 99.5\% Value at Risk (VaR). The solvency capital requirement makes our model similar to capacity constrained models, which are widely investigated in the literature. The similarity is straightforward but there are also important differences. Firstly, in case of the insurance industry there is no 'physical' limit on selling insurance contracts.

In product markets, a firm cannot sell more product than what is produced. An insurance company can sell more policies than what the solvency capital allows. Second, the 'capacity constraint' depends on premiums; at a higher level of premium greater number of policies can be sold (without violating the solvency capital requirement).

Solvency II criteria have an extensive literature \citep{Doff}. Some related papers deal with the question of portfolio optimization and asset allocation under Solvency II \citep{Kouwenberg, Escobar}, and some study the investment and reinsurance strategies under VaR constraints \citep{Bi, Zhang}. Although the literature on solvency regulations is abundant, the impact of the VaR constraint on the market equilibrium is rarely studied. \green{\citet{Dutang} built a non-cooperatvive game to study non-life insurer companies  market premium, solvency level, market share and underwriting results. \citet{Mouminoux} improve a similar repeated game and determine long run market shares, leadership and ruin probabilities and the effect of deviaton from the regulated market.}

The contribution of this paper bears on the effect of the solvency capital requirement. \green{Firstly, we assume oligopol market, where the companies decide on prices assuming fix level of same level of capital.} Introducing the solvency capital requirement constraint leads to the existence of a continuum of symmetric Nash equilibrium premiums, and in some cases the companies can achieve positive profits. This can occur, when the interest rate is low enough. If the number of companies with the same level of capital is increasing, then the set of equilibrium prices entails lower prices. If the capital of each firm is increasing, then the upper, and in some cases the lower endpoint of the Nash equilibrium interval decreases. Higher level of capital can lead to lower equilibrium premiums. If the total capital level is fixed in the market, the higher number of companies leads to lower level of individual capital leading to higher possible equilibrium premiums in the sets. In most cases, allowing mergers leads to a lower equilibrium premium. Even one company with higher level of capital (in a monopolistic situation) can ensure lower premium than what is achievable in an oligopoly market. This covers the case of a natural monopoly where only one company can operate efficiently in the market due to high fixed costs or barriers to entry. Railways and telecommunications are good examples.\green{ We also determine equlibrium prices in a duopoly market assuming different level of capital.}

\green{In the second part of the paper the companies can decide on level of the capital. We use two different approaches, first, after the price decision, companies can supplement the capital level to the necessary extent. In the other case, the companies make a simultaneous capital decision first and then a simultaneous price decision. Many market anomalies remain valid even after the endogenization of the capital decision, there may be a positive expected profit, one company dominates the market, or even market failure.}

\green{The paper is organized as follows.  Section \ref{sec:mod} presents the basic model.  In Section \ref{sec:short}, we describe the homogenous market equilibrium and the effect of changes in the number of insurance companies and the capital in the short run, and the market equlibrium strategies assuming heterogen capital level in a duopoly market. In the short run, the level of solvency capital is considered fixed, the companies cannot change it. In Section \ref{sec:endog} companies can change the level of their solvency capital, in section \ref{sec:expost} they can adjust some extra capital, if it is neccesary, and in section \ref{sec:exante} we assume a two period game (capital and price decision).} Finally, Section  \ref{sec:conc} concludes.

\section{Model}\label{sec:mod}

We model the situation as a non-cooperative game. In this game the set of players  $I$ are the insurance companies (\(i=1,\dots,I\)). We assume Bertrand premium competition, companies (players)  decide on the premium level (\(P_i\)) simultaneously. 

In our model the customers are homogeneous with respect to risk, but their reservation prices are different. Each  customer incurs a financial  loss \(K\) ($K>0$) with probability \(q\) ($0<q<1$), furthermore we assume that claims are independent. The risks of the customers are independent random variables $K \cdot  \kappa_j$, where \(\kappa_j\) is a random variable with Bernoulli distribution with parameter $q$. Customers can cover their losses by buying an insurance policy. If someone buys an insurance policy and financial loss occurs, then the insurance company will cover it completely (full coverage). Insurance contracts are homogeneous products, thus customers are indifferent between buying policies from any of the insurers.

Customers' intentions to buy are different, they are represented by a demand function \(D(P)\), which shows how many people buy insurances at premium $P$. The most simple --and most usual-- case, if we assume linear demand function (see for instance \citet{MasColelletal}, in insurance context \cite{KligerLevikson}). Unfortunately, in our case the linear demand function would lead to
a quartic expression, which would result very cumbersome formulas. We specify the demand curve as \(D(P)=\frac{\alpha^2}{P^2}\) (, if $qK\leq P$ and $\alpha>0$). This function belongs to the family of iso-elastic demand curves which is also frequent both in theory (\citet{Tramontanaetal}) and practice (\citet{Huangetal}) and also in insurance models (\citet{Haoetal}). According to Mossin at net premium every consumers buy an insurance \citep{Mossin}, thus at net premium the demand reaches the maximum size, $N_{max}=\frac{\alpha ^2}{q^2K^2}$. The inverse demand function is \(D^{-1}(n)=\frac{\alpha}{\sqrt{n}}\) the number of policies sold $n$ is less than $N_{max}$. Insurance companies cannot choose between customers, at premium level \(P_i\) they have to serve all potential customers \footnote{In several member states insurance companies are obliged to serve their customers at the quoted premium \citep{directive}, this is important to avoid discrimination for instance in the case of health insurance and compulsory motor third party liability insurance. There is a time period of the year, when people can choose or change their insurance ('open enrollment') and the insurance company must insure them. Thus, we assume in the model, that companies cannot reject customers, they must serve all of them at the given premium. See also in \citet{Polborn}.}.

Denote \(n_i(P_i)\) the number of customers buying from insurer $i$ at premium \(P_i\). Customers buy insurance coverage from the cheapest available company. If it is not unique, i.e. more companies offer the same level of premium, each company gets equal share of the customers. \(P_{\min}\) denotes the smallest premium and \(\mathcal{M}\) is the minimum
set \(P_i=P_{\min}\), and \(|\mathcal{M}|\) stands for the cardinality of set \(\mathcal{M}\), showing the number of such insurers. \green{$P_{-i}$ is the price vector of the other companies, contains the prices of the other insurers' except $i$th.}
$$n_i(P_i, P_{-i})= \left\{
\begin{array}{cl}
	\frac{1}{|\mathcal{M}|} D(P_i) & \mbox{ if } i\in\mathcal{M} \\
	0  & \mbox{ if } i\notin\mathcal{M} \\
\end{array}
\right. $$

Because insurance is a risky business not everybody is eligible to be  part of it.  Every insurance company faces a solvency capital requirement, in other words, they have to have enough capital to cover unexpected losses. The main concept of  Solvency II framework is that any insurance company can go bankrupt no more than once in every 200 cases. From a mathematical viewpoint the solvency capital requirement is a one-year 99.5\% Value at Risk (VaR) constraint.

\(\mbox{VaR}_\beta\) is a risk measure, which shows the maximum level of loss over a given time period, at a given confidence level. If $X$ is a continuous random variable with distribution function $F(.)$, then \(\mbox{VaR}_{\beta}(X)=\inf\{x | F(x)\geq \beta\} \) at confidence level $\beta$ i.e. insurance company $i$
fulfills the solvency capital requirement if its capital \(C\) plus premium incomes (\(P_i\) per policy) cover losses with probability 0.995:
\begin{equation} \label{eq:cap_solv}
	\mbox{Prob}\left(\sum_{j=1}^{n}K\kappa_j>C+n P_i\right) < 0.005 \ ,
\end{equation}
where \(n\) is an integer, and stands for the number of policies the insurance company has. If \(n\) is sufficiently large, then the distribution of the sum \(\sum_{j=1}^{n}K\kappa_j\) can be approximated by a normal distribution with mean \(nqK\) and standard deviation \(\sqrt{nq(1-q)}K\). Using the normal distribution approximation, equation (\ref{eq:cap_solv}) can be reformulated in a way that for given \(P_i\) and \(n\) the minimum capital requirement (\(MCR\)) is:
\begin{equation} \label{eq:cap_solv2}
	MCR(n,P_i)=n(qK-P_i)+\sqrt{n}\phi\sqrt{q(1-q)}K \ ,
\end{equation}
where \(\phi=\Phi^{-1}(0.995)\) and \(\Phi(x)\) is the cumulative distribution function of the standard normal distribution.

Let us suppose that company \(i\) has capital \(C\), which is a given parameter, not a decision variable. Holding capital has some costs, because the insurance company loses the interest \(rC\) on this capital. This is a fixed cost, it must be paid even if the insurer does not sell any contracts. Without this capital, the insurer would not be allowed to enter the market.

\begin{remark} \label{re:decreasing_ret_scale}
It is easy to see, that serving out twice as many customer requires less than twice as many capital at the same premium level.
In general way increase of the number of customer to \((1+a)n\) raises the solvency capital requirement less than to \((1+a)MCR(n,P_i)\) (,where a>0), minimum capital requirement has a decreasing return to scale in \(n\).
\end{remark}

The companies' payoff is their expected profit. The expected profit of company \(i\) is \(n_i(P_i)(P_i-qK) - r C\), if it fulfills the solvency capital requirement (i.e. \(MCR(n(P_i),P_i)\le C\)). If it is not satisfied, the company faces a penalty \(A\), which
is partly due to a financial penalty levied by the insurance supervisor, and to the decreased reputation of the firm.\footnote{If the insurer fails to meet the solvency capital requirement or is at risk of doing so, the insurer shall immediately notify the supervisor authority and submit a realistic recovery plan within two months. The supervisor authority shall require the insurer to make up the level of the solvency capital requirement or to reduce its risk profile to ensure compliance with the solvency capital requirement. If according to the supervisor authority the financial situation of the obligation concerned will deteriorate further, it can also restrict or
	prohibit the free disposal of the assets of that liability (DIRECTIVE 2009/138/EC, 2009).} The penalty is so great that selling no policies is preferred to selling many (profitable) with penalty. The expected profit of company $i$ can be written as:
\[\pi_i(P_i)
\left\{
\begin{array}{ll}
	\green{n_i(P_i, P_{-i})}(P_i-qK) - r C \ , & \mbox{ if }\hspace{1em} MCR(\green{n_i(P_i, P_{-i})},P_i) \le C \\
	\green{n_i(P_i, P_{-i})}(P_i-qK) - r C-A \ , \hspace{1em} & \mbox{ if }\hspace{1em} MCR(\green{n_i(P_i, P_{-i})},P_i) > C \\
\end{array}
\right.
\]

The companies' payoff is the expected profit, which we briefly refer to as profit from now on. Referring to Remark \ref{re:decreasing_ret_scale} we can state, that twice as large company's profit is more than twice as large at the same premium level.

Part of the profit is the technical result which is the profit without the interest loss and financial penalty:  \(TR_i=\green{n_i(P_i, P_{-i})}(P_i-qK)\). We show that insurance
companies' short term behaviour depends on the technical result.

Equation (\ref{eq:cap_solv}) can be rearranged in a way that it shows for given \(n\) and \(C\) the minimum premium levels needed to fulfill the solvency capital requirement. It is the minimum premium requirement
\begin{equation} \label{eq:cap_solv3}
	\mbox{MPR}(n,C)=qK-\frac{C}{n}+\frac{\phi\sqrt{q(1-q)}K}{\sqrt{n}} \ .
\end{equation}

The first part of equation(\ref{eq:cap_solv3}) can be interpreted as the net premium, this is the reference point. The second part is an increasing term in $n$, meaning that the more policies are taken by the company, the less portion of the capital can be allocated to a single policy and therefore the insurance company becomes riskier (which can be covered by an increased premium level)\footnote{If the company has big enough capital (and small number of policies) even negative premiums could be determined, but it is only a technical mathematical assumption.}. The third part is a decreasing term meaning that even if the variance is increasing for the whole portfolio but not in a linear way, less risk margin is needed for a greater portfolio (in a heuristic ---but not precise--- interpretation we can say that this is the positive effect of the law of large numbers). Formula (\ref{eq:cap_solv3}) can determine a higher level of MPR than $K$ under certain parameters. This level of the MPR is meaningless, because it is higher than the maximal level of claim. This is due to the inaccuracy of the approximation with the normal distribution, for meaningful parameters we get lower levels of MPR than $K$, see Appendix A.

The greatest premium (for a given capital) can be calculated, if we maximize $MPR(n, C)$ in $n$. The premium is the highest when the company has a large enough number of policies to share the capital amongst many policies and not large enough to take advantage of \lq the law of large numbers'. The maximum is reached at \(n=\frac{4C^2}{\phi^2q(1-q)K^2}\), while the maximal premium is:
\begin{equation}
	qK+\frac{\phi^2q(1-q)K^2}{4C}.
\end{equation}
It is easy to see that for smaller values than $\frac{4C^2}{\phi^2q(1-q)K^2}$ the MPR curve is increasing in $n$, and for higher values of $n$ it is decreasing. It is also obvious from expression (\ref{eq:cap_solv3}) that for a fixed capital the MPR function
tends to \(qK\) as \(n\) tends to infinity.

\begin{prop}\label{prop:inc_prof}
Assuming a fixed capital if $n_1< n_2$, than $ \pi_i\bigg(MPR(n_1,C),n_1\bigg) < \pi_i\bigg(MPR(n_2,C),n_2\bigg)$, i.e. the expected profits of insurers are increasing along the MPR function.
\end{prop}

\begin{proof}
Using expression (\ref{eq:cap_solv3}) for $MPR(n, C)$, the profit can be written:

\[
\pi_i\bigg(MPR(n,C),n\bigg)=\bigg(MPR(n,C)-qK\bigg)n - rC=
\]

\[
\left(qK-\frac{C}{n}+\frac{\phi\sqrt{q(1-q)}K}{\sqrt{n}}-qK\right)n-rC
=-(1+r) C+\sqrt{n}\phi\sqrt{q(1-q)}K
\]
The expected profit is increasing in $n$. 
\end{proof}

\begin{figure}[ht!]
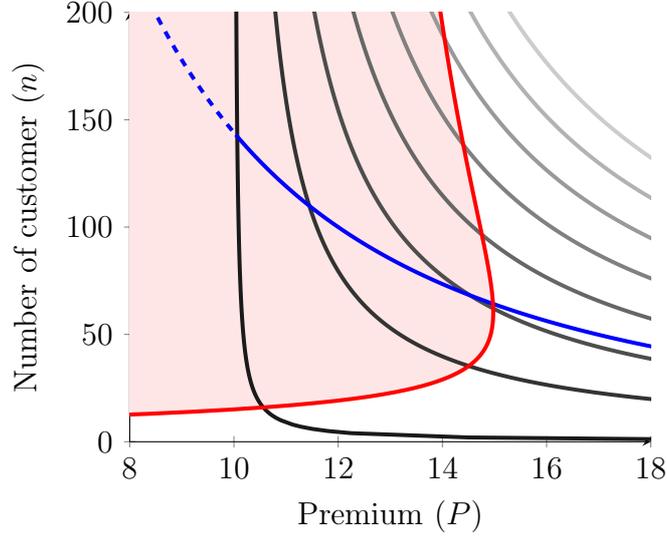

	\centering
	\pgfplotsset{scaled y ticks=false}
	\pgfplotsset{every axis plot/.append style={line width=1.5pt}}

	
	\caption{Illustration of the demand function, the isoprofit curves and the MPR capital requirement constraint in the ($P$, $n$) plane. $q=0.1, K=100, C=300, r=3\%, \alpha =120$}
	\label{fig:var_gen}
\end{figure}

We put the isoprofit, demand and MPR curves into a common plot as can be seen in Figure  \ref{fig:var_gen}. The plot is more readable if the horizontal axis is the premium and the vertical axis is the number of policies. The MPR function does not have an inverse function. We refer to it as MPR curve (red line). The shaded pink area is not available for companies, these pairs of premium and number of policies do not fulfill the solvency requirements. So companies can compete in the area right to the red curve. The gray curves are isoprofit curves, the lighter the curve the greater the expected profit.

In the next sections we analyze the possible market situation, and we show that this type of market situation leads to many anomalies. 

\begin{prop}\label{prop:intersection}
	The inverse demand function and the MPR function (for a fixed capital) has exactly one intersection point on interval \((0,\infty)\) at
	\[
	n=\frac{\left(-(\phi\sqrt{q(1-q)}K-\alpha)+\sqrt{(\phi\sqrt{q(1-q)}K-\alpha)^2+4qKC}\right)^2}{4q^2K^2} \ ,
	\]
and the premium at the point of intersection is
	\begin{equation} \label{eq:p_U}
	\frac{2\alpha q K}{-(\phi\sqrt{q(1-q)}K-\alpha)+\sqrt{(\phi\sqrt{q(1-q)}K-\alpha)^2+4qKC}} \ .
	\end{equation}
\end{prop}
 \green{This is the lowest premium level such that an insurer can cover the whole market alone withouth penalty.}

\begin{proof}
	See in Appendix B .
\end{proof}

\begin{remark}
		The premium at the intersection point is higher than \(qK\) if
	\[
	\frac{\alpha}{C}\phi\sqrt{\frac{1}{q}-1}> 1 \ .
	\]

	If the premium at the intersection point is less than \(qK\) the solvency capital requirement does not have an important role in the market, companies
	compete as if there were no capital requirements. 
\end{remark}

\begin{remark}
The point of intersection is left to the maximum of the MPR function (or in other words it is on the increasing part of the MPR curve)
if the following condition holds:
\begin{equation} \label{eq:inc_part_cond}
1>\frac{\phi^2(1-q)\left(-(\phi\sqrt{q(1-q)}K-\alpha)+\sqrt{(\phi\sqrt{q(1-q)}K-\alpha)^2+4qK\overline{C}}\right)^2}{16qC^2} \ .
\end{equation}
\end{remark}

The fact that the point of intersection is on the increasing or the decreasing part of the MPR curve plays an essential role in determining the type of short time equilibrium.

\section{Short term equilibrium in the insurance market with solvency capital requirement} \label{sec:short}

In this section we investigate the insurance market with solvency capital requirement. For a benchmark we consider the case without solvency capital requirement. In this case in a Bertrand oligopoly game every company sets premium $qK$ and the expected profit is zero for all the insurance companies. This phenomenon is usually called Bertrand paradox. Additional assumptions make the paradox disappear  (ensuring positive profits for firms in equilibrium). In the context of insurance markets see \citet{Polborn}. We will show that by adding solvency capital requirement positive and negative profits are also possible. 

We investigate a model, where companies have the same level of capital, \(C, \forall i\). Before offering insurance contracts
the companies had determined the level of solvency capital, so the interest on it \(rC\) is a sunk cost. In in other
words they suffer the interest loss even if they have no contract at all. The companies will sell insurance products as long as their technical result \(\green{n_i(P_i, P_{-i})}(P_i-qK)\) is nonnegative.

\subsection{Increasing part of the MPR curve is relevant in the market} \label{sec:short_inc}

First we consider the case when the increasing part of the MPR curve is relevant, the point of intersection is on the increasing part of the MPR curve. Expression (\ref{eq:inc_part_cond}) gives a condition for it.
The point of intersection of the demand function and the MPR curve is denoted with \(P_U\) and defined by (\ref{eq:p_U}). The \(I\)th part of the demand is \(D_I(P)=\frac{D(P)}{I}=\frac{\alpha^2}{I P^2}\), its inverse is\(D^{-1}_I(n)=\frac{\alpha}{\sqrt{I}\sqrt{n}}\). Thus, it can be considered as an inverse demand function with a modified parameter
\(\alpha'=\frac{\alpha}{\sqrt{I}}\). Therefore the premium \(P_L\) at the point of intersection of \(D_I(n)\) and the \(MPR\) function is, \green{which lowest premium such that an insurer who serves 1/I of the market fulfils the capital requirement.} 
\begin{equation} \label{eq:P_L}
P_L=\frac{2\frac{\alpha}{\sqrt{I}} qK}{-(\phi\sqrt{q(1-q)}K-\frac{\alpha}{\sqrt{I}})+\sqrt{(\phi\sqrt{q(1-q)}K-\frac{\alpha}{\sqrt{I}})^2+4qKC}}
\end{equation}

\begin{prop} \label{prop:NE}
If \(P_U>qK\), then there exists a continuum of symmetric Nash equilibria in the interval \([\max(qK,P_L),P_U]\).
\end{prop}

\begin{proof}

Since the inverse demand function intersects the MPR curve in its increasing part we can state that \(P_L < P_U\).

Let us suppose that all companies set premium level \(P_E\in[\max(qK,P_L),P_U]\). None of the companies intends to deviate
from the equilibrium level. If \(P_E\in[\max(qK,P_L),P_U]\), the technical result is nonnegative. Setting a higher premium results in no customers and so the technical results are 0. However, setting a lower premium means that the insurance company should serve the whole market
alone, but in the premium interval covering the whole market, does not satisfy the solvency capital requirement. Thus the company should pay a penalty, meaning that the profit is $\green{n_i(P_i, P_{-i})}(P_i-qK) - r C-A$, which is worse than selling no contract at all according to our initial assumption.

Premiums higher than $P_U$ cannot be equilibria, because in this case decreasing the premium and covering the whole market would lead to a higher level of expected profits.

Premiums lower than $qK$ cannot be Nash equilibria, since in this case the technical result is negative, setting higher
premium would mean zero technical result. Similarly, lower premium than $P_L$ cannot be a Nash equilibrium point, since all of the companies
face a penalty, and by setting a higher premium this penalty could be avoided.
\end{proof}

\begin{figure}[ht!]
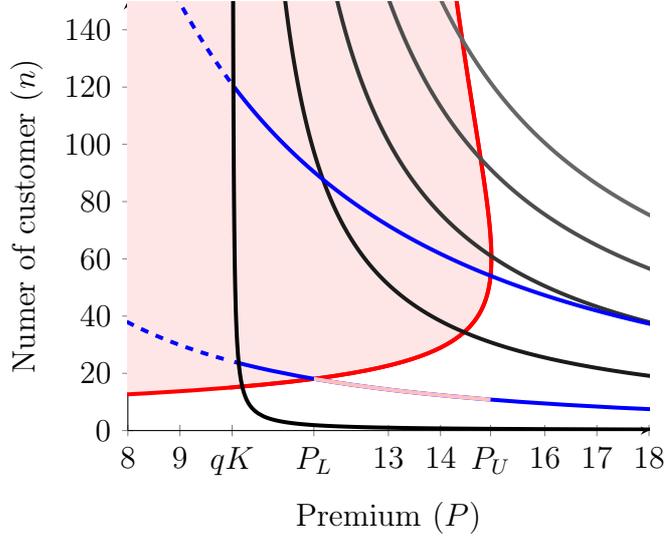

	\centering
	\pgfplotsset{scaled y ticks=false}
	\pgfplotsset{every axis plot/.append style={line width=1.5pt}}

	
	\caption{$q=0.1, K=100, C=300, r=1\%, \alpha=110, I=5$}
	\label{fig_eq_hom}

\end{figure}

An illustration of a continuum of Nash equilibria can be found in Figure \ref{fig_eq_hom}.
The red line is the solvency capital requirement, the blue lines are the demand functions and the $I$th part of them. The gray lines are the isoprofit curves. $P_L$ is the intersection point of the MPR curve and the $I$th part of the demand function, and $P_U$ is the intersection of the MPR and the demand function. Every premium between $P_L$ and $P_U$ can form an equilibrium. In these cases the market is somewhere on the pink line. The $I$ companies share the market equally, and they satisfy the solvency capital requirement.

In this concrete case the lowest possible equilibrium premium is $P_L$ is higher than the net premium. This is the lowest equilibrium premium most customers buy an insurance at. The highest possible equilibrium premium is $P_U$. At that premium even one company could serve the whole market without the possibility of paying a penalty.

\begin{figure}[h!]
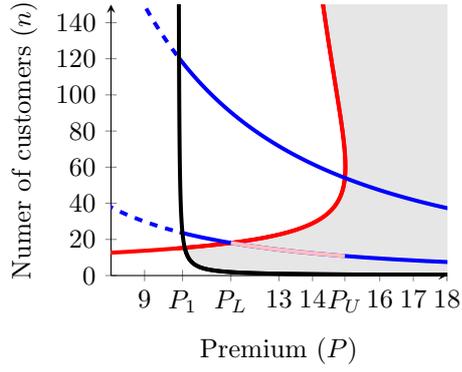
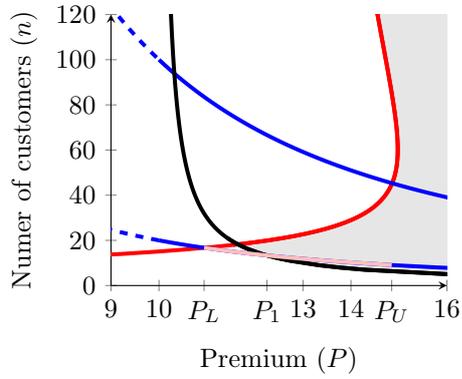
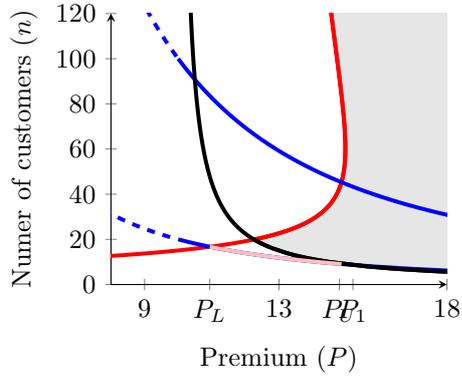

	\footnotesize
	\centering
	\pgfplotsset{scaled y ticks=false}
	\pgfplotsset{every axis plot/.append style={line width=1.5pt}}
	\begin{subfigure}{.5\textwidth}

		
		\caption{$q=0.1, K=100, C=300, r=1\%, \alpha=110, I=5$}
		
	\end{subfigure}

	
	\begin{subfigure}{.5\textwidth}
		%
		\caption{$q=0.1, K=100, C=300, r=10\%, \alpha=100, I=5$}

	\end{subfigure}
	
	\begin{subfigure}{.5\textwidth}
		%
		\caption{$q=0.1, K=100, C=300, r=15\%, \alpha=100, I=5$}
		
	\end{subfigure}
	\caption{Illustration of the cases of positive expected profit.}
	\label{fig_extraprofit}
\end{figure}

An important question is the profit of the insurance companies. We see that the technical result is nonnegative in every Nash equilibrium point, but the companies incur loss of interest as well. In Figure \ref{fig_extraprofit} the solid black curves are the zero-profit curves. Left to it the profit is negative, right to it is positive. It is easy to see that if the interest rate is high enough, then all Nash equilibria give negative expected profits, see the c) part in Figure \ref{fig_extraprofit}. On the other hand if \(P_L > qK\) and interest rate is small enough, then all Nash equilibria give positive profit (extraprofit), see part a) in Figure \ref{fig_extraprofit}. It can happen that some Nash equilibria give positive profit while others give negative profit, see part b) in Figure \ref{fig_extraprofit}.

In short, an equilibrium may exist where the companies have negative profits, because by selling some contracts they can reduce the loss of capital. But in the long run, the negative profit means that the sector is not profitable. On the other hand, the possibility of extra profit draws further companies to the market.

\begin{remark}
Considering the definition of the MPR function (expression (\ref{eq:cap_solv3})) it is quite clear that a decrease in the value of \(\phi\)
moves the function downward which means that the MPR curve moves to the left in Figure \ref{fig_eq_hom} which results in lower \(P_U\) and \(P_L\)
values. So there is a trade-off between safety and equilibrium price.
\end{remark}

Usually one of the most important issues in oligopoly model is the equilibrium price. As wee see in Proposition \ref{prop:NE} the equilibrium premium (premium)
is somewhere in interval \([\max(qK,P_L),P_U]\). An important question is, how the competition (increasing the number of insurance companies) affects to
this interval.

\begin{prop} \label{prop:incr_comp}
For a higher number of companies \(P_U\) remains unchanged while \(P_L\) decreases.
\end{prop}

\begin{proof}
\(P_U\) remains unchanged, since this is the (lowest) premium at which an insurance company
can insure the whole market. This can be seen in expression (\ref{eq:p_U}) which does not contain variable \(I\).

Decrease of \(P_L\) is quite trivial. Increasing the number of insurance companies leaves the MPR curve unchanged, while the $I$th part of the inverse demand curve moves downward. Since we are in the increasing part of the MPR curve, the point of intersection has to move to the left, which means lower premium.

Crucial in the argument is that we are in the increasing part of the MPR curve, in the decreasing part exactly the opposite is true. We have omitted the algebraic proof because it is too technical. 
\end{proof}

According to Proposition \ref{prop:incr_comp} the set of equilibrium prices entails lower prices as the number of insurance companies increases,
which agrees with economic intuition. However, in Proposition \ref{prop:incr_comp} as the number of
companies increases, the aggregate level of solvency capital in increasis. One can ask whether the increased number of companies alone has a reduction effect. To put it differently, what is the effect of increasing the number of companies if  the aggregate level of the solvency capital is fixed?

\begin{prop} \label{prop:incr_comp_fixC}
Assuming fixed total capital level in the market (each firm have capital level $\frac{C}{I}$) by increasing the number of companies both $P_U$ and $P_L$ increase.
\end{prop}

\begin{proof} 
For a fixed level of aggregate capital, definition of \(P_U\) changes to:
\[
P^C_U(I)=\frac{2\alpha q K}{-(\phi\sqrt{q(1-q)}K-\alpha)+\sqrt{(\phi\sqrt{q(1-q)}K-\alpha)^2+4qK\frac{C}{I}}} \ ,
\]
which is an increasing function in \(I\).

Similarly, the definition of \(P_L\) changes to:
\[
P^C_L(I)=\frac{2\frac{\alpha}{\sqrt{I}} q K}{-(\phi\sqrt{q(1-q)}K-\frac{\alpha}{\sqrt{I}})+\sqrt{(\phi\sqrt{q(1-q)}K-\frac{\alpha}{\sqrt{I}})^2+4qK\frac{C}{I}}} \ .
\]

With some algebra we get that \(P^C_L(I)\) is also an increasing function of $I$, see Appendix C.

In this situation, both the MPR curve and the \(I\)th part of inverse demand function change. With
graphical analysis alone we cannot prove the statement. It is interesting that in this case the algebraic proof is
the simpler, and we do not need condition (\ref{prop:incr_comp}). This statement is true generally (i.e. in the decreasing
part of the MPR curve as well).
\end{proof}

The result of Proposition \ref{prop:incr_comp_fixC} is quite interesting: the more concentrated the market is the lower is the
equilibrium premium (premium). This result is quite unusual in economic models. It suggests that a ban of mergers  of big companies would be disadvantageous for customers. Going further with this idea: is a monopoly market better for customers than an oligopoly market?

For the answer we need some more definitions. In case of monopoly \(P^C_L(1)=P^C_U(1)=P_1\). Since a monopolist does not face competition
 can set higher premium than \(P_1\). Let \(P_M\) be the premium which ensures the highest technical result for the insurance company (without solvency capital requirement).
By doing a little algebra one can derive that premium \(P_M=2qK\) will give the highest technical result. So the monopoly would like
to set premium \(P_M\), but it is not possible if \(P_M<P_1\). The monopoly sets premium \(\max(P_1,P_M)\).

\begin{figure}[ht!]
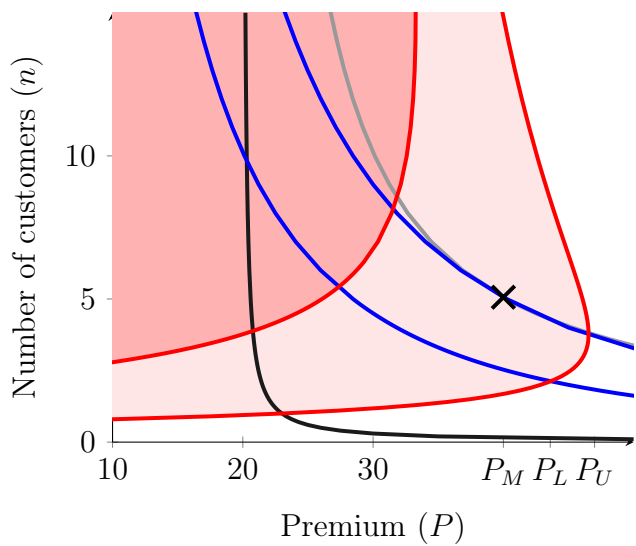

	\centering
	\pgfplotsset{scaled y ticks=false}
	\pgfplotsset{every axis plot/.append style={line width=1.5pt}}

	
	\caption{ $q=0.2, K=100, C=100, r=3\%, \alpha =90, I=2$}
	\label{fig:monopol}
\end{figure}

\begin{corol} \label{corol:monopoly}
If \(qK <P^C_L(2)\)
and \(P_M < P^C_L(2)\), then a monopoly determines a premium lower than any insurance companies in an oligopoly market (in Figure \ref{fig:monopol} we demonstrate such a situation).
\end{corol}

\begin{remark}
Proposition \ref{corol:monopoly} states that a monopoly market can produce a lower equilibrium premium than an oligopoly
market, which is advantageous for customers. It is not hard to deduce but worth to mention that (aggregate) profit also could be higher for a monopoly than for oligopoly firms.
\end{remark}

\subsection{Decreasing part of the MPR curve is relevant in the market} \label{sec:short_dec}

In this section we investigate the case, when the point of intersection is on the decreasing part of the MPR curve. We will see,
that in this case different equilibria can appear.

\begin{remark} \label{rem:decr_PL}
The fact that the point of intersection of the MPR curve and the inverse demand curve is on the increasing part of the MPR curve
does not necessarily mean that the point of intersection of the MPR curve and the \(I\)th part of the inverse demand curve is on the decreasing
part of the MPR curve either. On the contrary, if we look at expression (\ref{eq:P_L}), value \(P_L\) tends to zero as the number
of companies (\(I\)) tends to infinity. Therefore we have to be in the increasing part of the MPR curve because we know that the
limit of the MPR function at infinity is \(qK\).
\end{remark}

\begin{remark}
Following the idea in Remark \ref{rem:decr_PL}, for large enough \(I\) there is a continuum of Nash equilibrium points, since
\(P_L<P_U\). 
\end{remark}

\begin{prop} \label{prop:decrNE}
Consider an oligopoly market with \(I\) companies and suppose that in this market \(P_L(I)>P_U(I)\). If \(P_M\le P_U(I)\). Then there is only one type of Nash equilibrium point: one arbitrary company sets premium \(P_U\), the others set higher
premium.
\end{prop}

\begin{proof}
Let us suppose that companies set premiums \(P_1,\dots P_I\), the minimum of its is \(P_{\min}\). There are two cases:

a) \(P_{\min} > P_U\), in this case it is advantageous for any company to set premium a bit lower than \(P_{\min}\). At this premium the company gets the whole market ensuring higher profit than before.

b) \(P_{\min} = P_U\) and more than one company sets this premium. Than it causes loss for them since they cannot fulfill the
solvency capital requirement. If only one company sets premium \(P_U\) and \(P_M\le P_U\), then this is a Nash equilibrium. If
\(P_M>P_U\) than this cannot be a Nash equilibrium, since setting a premium a little bit higher than the minimum would increase the company's profit.
\end{proof}

\begin{remark}
If there are finitely-many possible premiums (\(P_U=P_1 < P_2 < \cdots P_F\)), then almost the same strategy is a Nash equilibrium as before:
one of the companies sets premium \(P_U\), and one of the others sets premium \(P_2\). The proof is the same as the proof of Proposition
\ref{prop:decrNE}. One problematic situation can arise, if profit at premium \(P_2\) is much more higher than profit at premium \(P_1\). In this situation it can happen for instance that half of the market at premium \(P_2\) gets greater profit than the whole market at premium \(P_1\). It is unrealistic if possible premium values are quite close to each other. The proof does not work if countable- or continuum-many premiums are possible, but we feel that this case is far from reality.
\end{remark}

We know from Proposition \ref{prop:incr_comp_fixC} that for a fixed level of aggregate capital the increased number of
insurance companies increases both \(P_U\) and \(P_L\) values. So again, low equilibrium premium can be achievable if few big
companies are in the market and the merger of companies (even big companies)  is advantageous for customers in many occasions.

We can also state that monopoly can set a lower premium than competitive companies in an oligopoly situation.
The condition changes a bit: \(P_M<\min(P^C_L(2),P^C_U(2))\).

\subsection{Asymmetric capital level}

\begin{figure}[ht!]
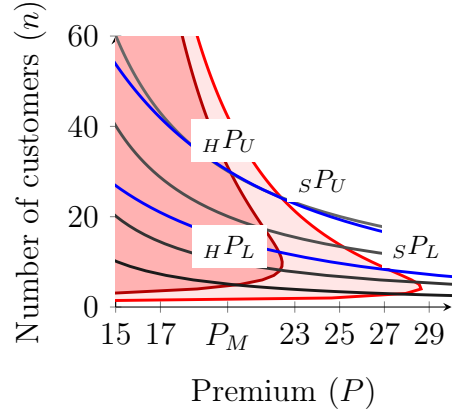

	\centering
	\pgfplotsset{scaled y ticks=false}
	\pgfplotsset{every axis plot/.append style={line width=1.1pt}}

	
	\caption{\footnotesize{$q=0.1, K=100, {_S}C=80, {_H}C=120, r=1\%, \alpha=110, I=2$}}
	\label{fig_eq_het_dec}

\end{figure}

\begin{figure}[ht!]
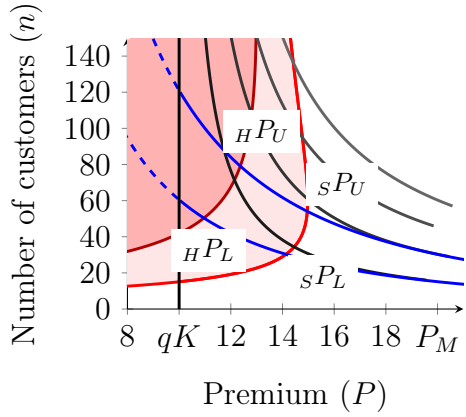

	\centering
	\pgfplotsset{scaled y ticks=false}
	\pgfplotsset{every axis plot/.append style={line width=1.1pt}}
	%
	
	\caption{\footnotesize{$q=0.1, K=100, {_S}C=300, {_H}C=500, r=1\%, \alpha=110, I=2$}}
	\label{fig_eq_het_inc}

\end{figure}

For investigating a model where the capital is endogenous we have to investigate the situation when insurance companies'
capital are different. For the sake of simplicity we describe only the situation in a two player market ($I=2$), the left index
refers to insurance companies' capital level: \(S\)=small, \(H\)= \green{high}. \red{Even in this frame quite a lot possibility may arose. We discribe two typical situation, all other cases is somehow a variation (or combination) of these two cases:}
\begin{itemize}
\item \red{Point of intersection of the demand function and the MPR curve is on the decreasing part of the MPR curve for both of the companies in a way that }\({_S}P_U < {_S}P_L\) \red{ and }\({_H}P_U < {_H}P_L\)\red{. We will refer to this situation as Case I. Case I is illustrated in part a) of Figure \ref{fig_eq_het_dec} .}
\item \red{Point of intersection of the demand function and the MPR curve is is on the increasing part of the MPR curve for both of the companies. For the sake of simplicity we also assume that} \(qK < {_H}P_L\) \red{which also includes} \(qK < {_S}P_L\).
\red{We will refer to it as Case II, and illustrated on Figure \ref{fig_eq_het_inc}.}
\end{itemize}

\red{First we investigate the Case I situation. We have to take more distinction in this situation: first we assume, that monopolistic premium }(\(P_M\))\red{\ is less than }\({_H}P_U \)
\red{\ (Case I/a), Case I/b will refer to when the monopolistic premium is between \({_H}P_U \) and \({_S}P_U\)
(\({_H}P_U\le P_M \le {_S}P_U\)). Case I/c when
the monopolistic premium is strictly higher than \({_S}P_U \). Case I/b and I/c can be investigated (or will be investigated) in a discrete setting,
the company can set premiums \({_H}P_U=P_1 < P_2 < \cdots <  P_J \). Index \(k\) refer to monopolistic premium (\(P_k= P_M\)), index \(\ell\) to \({_S}P_U\) .}

\begin{prop} \label{prop:caseIa}
\red{In Case I/a situation there exists single a pure Nash equilibrium: the large company set premium level }\({_H}P_U\)\red{\ the small company
set any premium level higher than this. The solution is single in a way that there exists no other pure Nash-equilibrium where
the large company sets higher premium level than \({_H}P_U\)}.
\end{prop}

\begin{proof} 
\red{The smaller company do not got any customer, until it sets higher premium level than }\({_H}P_U\)\red{. The smaller company
will not set premium level \({_H}P_U\) (or lower premium level) since in this case it face penalty \(A\), which is a worse situation. The larger company does not set lower premium level than }\({_H}P_U\)\red{\ since in this situation it would have to pay penalty \(A\). The larger company does not set higher premium also, since above the monopolistic premium any further increase would decrease the profit.}

\red{Let suppose that in a Nash-equilibrium the large company set premium level higher than \({_H}P_U\). If the small company set premium level higher than \({_H}P_U\) then it would be worth to undercut for the large company. If the small company sets
premium level \({_H}P_U\) it faces penalty \(A\). So the small company sets higher level of premium than the large company. But in this situation it is worth for the large company to decrease the premium level.}
\end{proof}

\begin{prop}
\red{In Case I/b situation there exists a single pure Nash equilibrium: the large company set premium level }\(P_M\)
\red{ the small company set premium level higher than it. This is single in a way that no pure Nash equilibrium exist where
the large company set higher or lower premium.}
\end{prop}

\begin{proof}
\red{It is quite easy to see that it is Nash-equilibrium, if large company set premium level \(P_M\) and the small company sets
higher level than this, the large company sets the monopolistic price, it does not want to deviate from it. The small company does not get any customer until it set higher premium level than \(P_M\) so it has no motivation deviate it until the premium remain
higher than \(P_M\). the small company cannot set lower or equal level of premium than \(P_M\), otherwise it would pay penalty \(A\).}

\red{Let us suppose that there exists another Nash equilibrium where the large company set a premium level which is different than
\(P_M\). The smaller company cannot sets a premium level which is lower or equal with \(P_M\), since at this premium it would face
penalty. If it would set premium level higher than \(P_M\) then it is worth for the large company to undercut. So the small company necessarily sets higher premium than the large company. But in this case it worth the large company to set a premium level which is
closer to \(P_M\) in a way that it is still lower than the small company's premium level.}
\end{proof}

\red{In proposition \ref{prop:caseIbdiscrete} we investigate the same situation in a discrete setting. In this frame there are
more than one equilibrium price.}

\begin{prop} \label{prop:caseIbdiscrete}
\red{In Case I/b situation there (may) exist more than one pure Nash equilibrium assuming finite many possible premium level: the large company set premium level \(P_{j}\) (\(j=1, \dots, k\)), the small company sets premium level \(P_{j+1}\).}
\end{prop}

\begin{proof}
\red{The large company is not interested in a lower premium level, so it will not decrease the premium level. If it increase, it will
get only half of the market which is a worse situation (we assume that getting the whole market at \(P_j\) is still better than getting the half of it at premium \(P_{j+1}\)). The small company's profit does not change if it sets higher level of premium, so it is
not intended to do this. If it decreases the premium level it will get so many customer that it will have to pay penalty.}
\end{proof}

\begin{prop} \label{prop:caseIccont}
\red{In Case I/c situation there exists no pure Nash equilibrium if companies can set any premium (continuously).}
\end{prop}

\begin{proof}
\red{Let us suppose that there exist a pure Nash equilibrium. The smaller company cannot set lower or equal premium than the large company
(see proof of Proposition \ref{prop:caseIa}). If the large company sets higher premium than \({_S}P_U\) then the small company will undercut. If large company would set premium \({_S}P_U\) and the small a higher one then the large company would raise premium (at least a bit).}
\end{proof}

\begin{prop}
\red{In Case I/c situation there (may) exist more than one pure Nash equilibrium: the large company set premium level \(P_{j}\)
(\(j=1, \dots, \ell\)), the small company sets premium level \(P_{j+1}\).}
\end{prop}

\begin{proof}
\red{The proof is essentially the same as proof of Proposition \ref{prop:caseIbdiscrete}.}
\end{proof}

\red{As we see, in Case I if there exist a pure Nash equilibrium the large company gets the whole market, however the discrete and continuous
setting may result different Nash-equilibrium price.}

\red{We move to investigate Case II. In Case II we again have to differentiate three cases. Case II/a when }\({_H}P_U<{_S}P_L\) and \(P_M<{_S}P_L\);\red{\ Case II/b when }\({_H}P_U<{_S}P_L\) and \green{\({_S}P_L \le P_M\)}\red{\ (Case II/a and II/b are represented part a and b in Figure \ref{fig:Cases}; Case II/c }
\green{\({_S}P_L\le{_H}P_U\)}\red{\ (part c in Figure \ref{fig:Cases}}).

\begin{prop}
\red{In Case II/a situation there exists a pure Nash equilibrium: the greater company set premium level \green{\(max({_H}P_U, P_M)\)}, the smaller company set any higher premium level.}
\end{prop}

\begin{proof}
\green{Essentially is the same as before in Proposition \ref{prop:caseIa}. The expected profit is increasing in $P$ under the monopoly premium level ($P_M$), and ${_H}P_U$ is the lowest premium at that the large company can serve the whole market without penalty. So it is worth to increase the premium until \(max({_H}P_U, P_M)\).}  
\end{proof}

\begin{prop}
\red{In Case II/b situation there exists no pure Nash equilibrium assuming continuous premium setting.}
\end{prop}

\begin{proof}
\red{Essentially is the same as proof of Propositon \ref{prop:caseIccont}.}
\end{proof}

\begin{prop}\label{prop:ACLIIcdD}
\red{In Case II/a and II/b situation there (may) exist more than one pure Nash equilibrium: the large company
set premium level \(P_j\) (\(j=1,\cdots \ell-1\)), the small company set premium \(P_{j+1}\).}
\end{prop}

\begin{proof}
\green{Essentially is the same as proof of Propositon \ref{prop:caseIbdiscrete}.}
\end{proof}

\begin{prop} \label{prop:ACLIIc}
\red{In Case II/c situation there exists countinuum-many pure Nash equilibrium: both the company set the same premium level in interval }\([{_S}P_L,{_H}P_U]\).
\end{prop}

\begin{proof}
\red{It is quite clear, that if both company set the same premium level in interval }\([{_S}P_L,{_H}P_U]\)\red{ is a Nash equilibrium: if any company undercut, it will
get the whole market, but face penalty }\(A\)\red{, which is a worse situation. On the other hand if any of the companies raise the premium, it will get no market share, which
is again a worse situation. The smaller company will not set premium lower premium than }\({_S}P_L\)\red{ since it cannot cover even half of the market. Higher
premium level than }\({_H}P_U\)\red{ also cannot be Nash equilibrium, since in this situation it is worth to undercut for the larger company.}
\end{proof}

\begin{figure}[h!]
	\footnotesize
	\centering
	\pgfplotsset{scaled y ticks=false}
	\pgfplotsset{every axis plot/.append style={line width=1.5pt}}
	\begin{subfigure}{.5\textwidth}
		\begin{tikzpicture}

			\begin{axis}[ 
				width=6cm,
				axis lines = left,
				xlabel = Premium ($P$),
				ylabel = Numer of customers ($n$),
				xmin=30, xmax=50,
				xtick={20,30,50},
				ymin=0, ymax=10,
				extra x ticks={37.01,40,43.63},
				extra x tick labels={${_H}P_U$, $P_M$, $_{_S}P_L$},
				ytick={0, 2,4,6,8},
				yticklabel style={
					/pgf/number format/fixed,
					/pgf/number format/precision=2
				},
				scaled y ticks=false,
				legend style ={ at={(1.03,1)}, 
					anchor=north west, 
					fill=white,align=left},
				]

				\addplot [
				color=red, 
				]
				coordinates {
					
				(23.033172141956,1)(42.855454708738,2)(46.152896338286,3)(46.516586070978,4)(46.0778353693702,5)(45.3964497213558,6)(44.6571643253891,7)(43.927727354369,8)(43.2332796028742,9)(42.581949852079,10)(41.9747611777242,11)(41.4097815024763,12)(40.8839527101177,13)(40.3939164019786,14)(39.9363839875145,15)(39.508293035489,16)(39.1068618637489,17)(38.7295960140238,18)(38.3742729052219,19)(38.0389176846851,20)(37.7217766872347,21)(37.4212915636014,22)(37.1360754306652,23)(36.8648915273446,24)(36.6066344283912,25)(36.3603136653538,26)(36.1250395201694,27)(35.9000107341231,28)(35.684503882286,29)(35.4778641844874,30)(35.2794975497198,31)(35.0888636771845,32)(34.9054700618486,33)(34.7288667745018,34)(34.5586419056471,35)(34.3944175792149,36)(34.2358464562934,37)(34.0826086611102,38)(33.9344090716706,39)(33.7909749260395,40)(33.6520537024817,41)(33.5174112377723,42)(33.3868300531326,43)(33.2601078615894,44)(33.1370562342345,45)(33.017499405976,46)(32.9012732040194,47)(32.7882240845715,48)(32.6782082651774,49)(32.5710909417476,50)(32.4667455807356,51)(32.3650532781358,52)(32.2659021780112,53)(32.1691869441556,54)(32.0748082792698,55)(31.9826724867036,56)(31.8926910703951,57)(31.8047803691475,58)(31.7188612218237,59)(31.6348586604239,60)(31.5527016283499,61)(31.4723227214545,62)(31.3936579497329,63)(31.3166465177445,64)(31.2412306220502,65)(31.1673552641321,66)(31.0949680774175,67)(31.0240191671686,68)(30.9544609621234,69)(30.8862480768827,70)(30.819337184135,71)(30.7536868959008,72)(30.6892576530523,73)(30.6260116224365,74)(30.5639126009905,75)(30.5029259262952,76)(30.4430183930611,77)(30.3841581750891,78)(30.3263147522842,79)(30.2694588423426,80)(30.2135623367605,81)(30.1585982408466,82)(30.1045406174437,83)(30.0513645340935,84)(29.9990460133962,85)(29.94756198634,86)(29.8968902483928,87)(29.8470094181643,88)(29.7978988984624,89)(29.7495388395819,90)(29.7019101046761,91)(29.6549942370717,92)(29.608773429401,93)(29.5632304944313,94)(29.5183488374836,95)(29.4741124303389,96)(29.4305057865372,97)(29.387513937983,98)(29.3451224127768,99)(29.3033172141956,100)(29.2620848007544,101)(29.2214120672823,102)(29.1812863269533,103)(29.1416952942154,104)(29.1026270685644,105)(29.0640701191153,106)(29.0260132699229,107)(28.9884456860106,108)(28.9513568600676,109)(28.9147365997747,110)(28.8785750157272,111)(28.8428625099187,112)(28.8075897647574,113)(28.7727477325849,114)(28.7383276256713,115)(28.7043209066602,116)(28.6707192794409,117)(28.6375146804247,118)(28.6046992702042,119)(28.572265425577,120)(28.5402057319134,121)(28.5085129758517,122)(28.4771801383051,123)(28.4462003877631,124)(28.415567073874,125)(28.3852737212945,126)(28.3553140237924,127)(28.3256818385922,128)(28.2963711809493,129)(28.2673762189436,130)(28.2386912684822,131)(28.2103107885,132)(28.1822293763504,133)(28.1544417633761,134)(28.126942810653,135)(28.0997275048979,136)(28.0727909545337,137)(28.0461283859036,138)(28.0197351396301,139)(27.9936066671093,140)(27.9677385271365,141)(27.9421263826576,142)(27.9167659976389,143)(27.8916532340519,144)(27.8667840489679,145)(27.8421544917571,146)(27.8177607013878,147)(27.7935989038224,148)(27.769665409505,149)(27.7459566109378,150)(27.7224689803415,151)(27.6991990673972,152)(27.6761434970666,153)(27.6532989674862,154)(27.6306622479332,155)(27.6082301768609,156)(27.5859996599998,157)(27.5639676685214,158)(27.5421312372648,159)(27.5204874630198,160)(27.4990335028677,161)(27.4777665725758,162)(27.4566839450442,163)(27.4357829488018,164)(27.4150609665519,165)(27.3945154337628,166)(27.3741438373039,167)(27.3539437141243,168)(27.3339126499729,169)(27.3140482781588,170)(27.2943482783488,171)(27.2748103754035,172)(27.2554323382478,173)(27.2362119787764,174)(27.2171471507921,175)(27.1982357489765,176)(27.1794757078909,177)(27.1608650010072,178)(27.1424016397678,179)(27.1240836726728,180)(27.1059091843939,181)(27.0878762949145,182)(27.0699831586941,183)(27.0522279638576,184)(27.0346089314069,185)(27.017124314456,186)(26.9997723974868,187)(26.9825514956267,188)(26.965459953946,189)(26.9484961467749,190)(26.9316584770401,191)(26.9149453756191,192)(26.8983553007123,193)(26.8818867372325,194)(26.8655381962115,195)(26.8493082142213,196)(26.8331953528129,197)(26.8171981979681,198)(26.801315359568,199)(26.7855454708738,200)(26.7698871880223,201)(26.7543391895349,202)(26.7389001758383,203)(26.7235688687992,204)(26.7083440112696,205)(26.6932243666453,206)(26.6782087184343,207)(26.6632958698371,208)(26.6484846433379,209)(26.6337738803051,210)(26.6191624406032,211)(26.6046492022132,212)(26.5902330608634,213)(26.5759129296689,214)(26.5616877387803,215)(26.5475564350408,216)(26.5335179816517,217)(26.5195713578466,218)(26.5057155585726,219)(26.4919495941804,220)(26.4782724901205,221)(26.4646832866478,222)(26.451181038533,223)(26.4377648147804,224)(26.4244336983526,225)(26.4111867859024,226)(26.3980231875097,227)(26.3849420264256,228)(26.3719424388221,229)(26.3590235735476,230)(26.3461845918877,231)(26.3334246673324,232)(26.3207429853475,233)(26.3081387431521,234)(26.295611149501,235)(26.2831594244712,236)(26.2707827992545,237)(26.2584805159538,238)(26.2462518273842,239)(26.2340959968786,240)(26.2220122980977,241)(26.2100000148439,242)(26.1980584408795,243)(26.1861868797487,244)(26.1743846446039,245)(26.1626510580351,246)(26.1509854519036,247)(26.1393871671788,248)(26.1278555537792,249)(26.1163899704158,250)(26.1049897844395,251)(26.0936543716918,252)(26.0823831163583,253)(26.0711754108254,254)(26.06003065554,255)(26.0489482588723,256)(26.0379276369811,257)(26.0269682136826,258)(26.0160694203205,259)(26.0052306956405,260)(25.9944514856662,261)(25.9837312435779,262)(25.9730694295939,263)(25.9624655108539,264)

				};

				\addplot [
				color=black!30!red, name path=VAR
				]
				coordinates {
					(-26.966827858044,1)(17.855454708738,2)(29.4862296716193,3)(34.016586070978,4)(36.0778353693702,5)(37.0631163880225,6)(37.514307182532,7)(37.677727354369,8)(37.6777240473187,9)(37.581949852079,10)(37.4293066322697,11)(37.2431148358097,12)(37.0377988639639,13)(36.8224878305501,14)(36.6030506541812,15)(36.383293035489,16)(36.1656853931607,17)(35.951818236246,18)(35.7426939578535,19)(35.5389176846851,20)(35.3408243062823,21)(35.1485642908741,22)(34.9621623871869,23)(34.7815581940112,24)(34.6066344283912,25)(34.4372367422769,26)(34.2731876683176,27)(34.1142964484088,28)(33.9603659512515,29)(33.8111975178207,30)(33.6665943239133,31)(33.5263636771845,32)(33.3903185466971,33)(33.2582785392077,34)(33.1300704770757,35)(33.005528690326,36)(32.8844951049421,37)(32.766819187426,38)(32.6523577896193,39)(32.5409749260395,40)(32.4325415073597,41)(32.3269350472961,42)(32.2240393554582,43)(32.1237442252258,44)(32.0259451231234,45)(31.9305428842369,46)(31.8374434167854,47)(31.7465574179048,48)(31.6578001019121,49)(31.5710909417476,50)(31.4863534238728,51)(31.4035148165973,52)(31.3225059515961,53)(31.2432610182297,54)(31.1657173701789,55)(31.0898153438465,56)(31.0154980879389,57)(30.9427114036303,58)(30.8714035947051,59)(30.8015253270906,60)(30.7330294972024,61)(30.6658711085512,62)(30.6000071560821,63)(30.5353965177445,64)(30.4719998528194,65)(30.4097795065563,66)(30.3486994207011,67)(30.2887250495215,68)(30.229823280964,69)(30.171962362597,70)(30.1151118320223,71)(30.0592424514563,72)(30.004326146203,73)(29.9503359467608,74)(29.8972459343239,75)(29.8450311894531,76)(29.7936677437105,77)(29.7431325340634,78)(29.6934033598791,79)(29.6444588423426,80)(29.5962783861433,81)(29.5488421432856,82)(29.5021309788895,83)(29.4561264388554,84)(29.4108107192785,85)(29.3661666375027,86)(29.3221776047146,87)(29.2788275999825,88)(29.2361011456534,89)(29.1939832840263,90)(29.1524595552255,91)(29.1115159762021,92)(29.0711390207988,93)(29.0313156008142,94)(28.99203304801,95)(28.9532790970056,96)(28.9150418690114,97)(28.8773098563503,98)(28.8400719077263,99)(28.8033172141956,100)(28.7670352958039,101)(28.7312159888509,102)(28.6958494337494,103)(28.6609260634461,104)(28.6264365923739,105)(28.5923720059078,106)(28.5587235502967,107)(28.5254827230477,108)(28.4926412637373,109)(28.4601911452293,110)(28.4281245652768,111)(28.3964339384901,112)(28.3651118886512,113)(28.3341512413568,114)(28.3035450169756,115)(28.2732864239016,116)(28.2433688520905,117)(28.2137858668654,118)(28.1845312029773,119)(28.1555987589104,120)(28.1269825914175,121)(28.0986769102779,122)(28.0706760732645,123)(28.0429745813115,124)(28.015567073874,125)(27.9884483244691,126)(27.9616132363908,127)(27.9350568385922,128)(27.9087742817245,129)(27.8827608343283,130)(27.8570118791693,131)(27.8315229097122,132)(27.8062895267263,133)(27.7813074350179,134)(27.7565724402826,135)(27.7320804460744,136)(27.707827450884,137)(27.6838095453239,138)(27.6600229094143,139)(27.6364638099664,140)(27.6131285980585,141)(27.5900137066013,142)(27.5671156479885,143)(27.5444310118297,144)(27.521956462761,145)(27.4996887383325,146)(27.477624646966,147)(27.4557610659845,148)(27.4340949397064,149)(27.4126232776045,150)(27.3913431525269,151)(27.3702516989761,152)(27.3493461114457,153)(27.3286236428108,154)(27.3080816027719,155)(27.2877173563481,156)(27.2675283224201,157)(27.2475119723189,158)(27.2276658284598,159)(27.2079874630198,160)(27.1884744966565,161)(27.1691245972672,162)(27.1499354787865,163)(27.1309049000213,164)(27.1120306635216,165)(27.0933106144857,166)(27.0747426396991,167)(27.0563246665052,168)(27.0380546618073,169)(27.0199306311,170)(27.0019506175301,171)(26.9841127009849,172)(26.9664149972073,173)(26.9488556569373,174)(26.9314328650778,175)(26.9141448398856,176)(26.8969898321847,177)(26.8799661246027,178)(26.8630720308293,179)(26.846305894895,180)(26.8296660904713,181)(26.8131510201892,182)(26.7967591149783,183)(26.7804888334228,184)(26.7643386611366,185)(26.7483071101549,186)(26.7323927183425,187)(26.7165940488182,188)(26.7009096893957,189)(26.685338252038,190)(26.6698783723281,191)(26.6545287089524,192)(26.6392879431993,193)(26.6241547784697,194)(26.6091279398012,195)(26.594206173405,196)(26.5793882462139,197)(26.5646729454429,198)(26.550059078161,199)(26.5355454708738,200)(26.5211309691169,201)(26.5068144370596,202)(26.4925947571191,203)(26.4784708295835,204)(26.4644415722452,205)(26.4505059200434,206)(26.4366628247145,207)(26.4229112544525,208)(26.4092501935771,209)(26.3956786422099,210)(26.3821956159586,211)(26.3688001456094,212)(26.3554912768258,213)(26.3422680698558,214)(26.3291295992454,215)(26.3160749535593,216)(26.3031032351079,217)(26.2902135596814,218)(26.2774050562895,219)(26.2646768669076,220)(26.252028146229,221)(26.2394580614226,222)(26.2269657918963,223)(26.2145505290661,224)(26.2022114761304,225)(26.1899478478493,226)(26.1777588703291,227)(26.1656437808116,228)(26.1536018274684,229)(26.1416322691998,230)(26.1297343754375,231)(26.1179074259531,232)(26.1061507106693,233)(26.0944635294769,234)(26.0828451920542,235)(26.0712950176915,236)(26.0598123351195,237)(26.0483964823404,238)(26.0370468064637,239)(26.0257626635453,240)(26.0145434184297,241)(26.003388444596,242)(25.9922971240071,243)(25.9812688469619,244)(25.9703030119509,245)(25.9593990255148,246)(25.948556302106,247)(25.937774263953,248)(25.9270523409278,249)(25.9163899704158,250)(25.9057865971885,251)(25.8952416732791,252)(25.8847546578603,253)(25.8743250171246,254)(25.8639522241674,255)(25.8536357588723,256)(25.8433751077983,257)(25.8331697640702,258)(25.8230192272703,259)(25.8129230033328,260)(25.8028806044401,261)(25.7928915489214,262)(25.7829553611528,263)(25.77307157146,264)
					
				};
				
				\addplot [
				color=blue, 
				]
				coordinates {
				(90,1)(63.6396103067893,2)(51.9615242270663,3)(45,4)(40.2492235949962,5)(36.7423461417477,6)(34.0168025708304,7)(31.8198051533946,8)(30,9)(28.4604989415154,10)(27.1360210119987,11)(25.9807621135332,12)(24.9615088301353,13)(24.0535117721182,14)(23.2379000772445,15)(22.5,16)(21.82820625327,17)(21.2132034355964,18)(20.6474160483506,19)(20.1246117974981,20)(19.6396101212393,21)(19.1880644720049,22)(18.7662972651367,23)(18.3711730708738,24)(18,25)(17.6504521624366,26)(17.3205080756888,27)(17.0084012854152,28)(16.7125804359347,29)(16.431676725155,30)(16.1644771824097,31)(15.9099025766973,32)(15.6669890360128,33)(15.4348726628258,34)(15.2127765851133,35)(15,36)(14.7959088574822,37)(14.5999279017686,38)(14.4115338424578,39)(14.2302494707577,40)(14.0556385699745,41)(13.8873014965883,42)(13.7248713299344,43)(13.5680105059994,44)(13.4164078649987,45)(13.2697760539407,46)(13.1278492348105,47)(12.9903810567666,48)(12.8571428571429,49)(12.7279220613579,50)(12.6025207562521,51)(12.4807544150677,52)(12.362450755382,53)(12.2474487139159,54)(12.1355975243384,55)(12.0267558860591,56)(11.9207912135854,57)(11.817578957375,58)(11.7170019882741,59)(11.6189500386223,60)(11.5233191939606,61)(11.4300114300171,62)(11.3389341902768,63)(11.25,64)(11.1631261130288,65)(11.0782341881399,66)(10.9952499920675,67)(10.914103126635,68)(10.8347267777192,69)(10.7570574840095,70)(10.6810349237447,71)(10.6066017177982,72)(10.5337032476518,73)(10.4622874869437,74)(10.3923048454133,75)(10.3237080241753,76)(10.2564518813674,77)(10.1904933073014,78)(10.1257911083342,79)(10.0623058987491,80)(10,81)(9.93883734673619,82)(9.87878339907213,83)(9.81980506061966,84)(9.76187060183953,85)(9.70494958830946,86)(9.64901281354015,87)(9.59403223600247,88)(9.53998092005724,89)(9.48683298050514,90)(9.43456353049726,91)(9.38314863256837,92)(9.33256525257383,93)(9.28279121632914,94)(9.23380516876639,95)(9.18558653543692,96)(9.13811548620257,97)(9.0913729009699,98)(9.04534033733291,99)(9,100)(8.9553347118899,101)(8.91132788679007,102)(8.86796350347864,103)(8.82522608121828,104)(8.7831006565368,105)(8.74157276121538,106)(8.70062840141097,107)(8.66025403784439,108)(8.62043656699036,109)(8.58116330321033,110)(8.54242196177249,111)(8.50420064270761,112)(8.46648781545237,113)(8.42927230423525,114)(8.39254327416282,115)(8.35629021796733,116)(8.32050294337844,117)(8.28517156108491,118)(8.2502864732539,119)(8.21583836257749,120)(8.18181818181818,121)(8.14821714382667,122)(8.11502671200689,123)(8.08223859120487,124)(8.04984471899924,125)(8.01783725737273,126)(7.98620858474502,127)(7.95495128834866,128)(7.92405815693061,129)(7.89352217376326,130)(7.86333650994934,131)(7.8334945180064,132)(7.80398972571708,133)(7.77481583023224,134)(7.74596669241483,135)(7.7174363314129,136)(7.68921891945085,137)(7.66130877682874,138)(7.63370036711974,139)(7.60638829255665,140)(7.57936728959867,141)(7.5526322246702,142)(7.52617809006382,143)(7.5,144)(7.4740931868366,145)(7.44845299742131,146)(7.4230748895809,147)(7.39795442874108,148)(7.37308728467136,149)(7.34846922834953,150)(7.32409612894044,151)(7.29996395088431,152)(7.27606875108999,153)(7.25240667622842,154)(7.22897396012249,155)(7.20576692122892,156)(7.1827819602086,157)(7.16001555758157,158)(7.1374642714633,159)(7.11512473537885,160)(7.09299365615191,161)(7.07106781186548,162)(7.04934404989162,163)(7.02781928498727,164)(7.00649049745371,165)(6.985354731357,166)(6.96440909280723,167)(6.94365074829414,168)(6.92307692307692,169)(6.90268489962633,170)(6.88247201611685,171)(6.86243566496721,172)(6.84257329142735,173)(6.82288239221013,174)(6.80336051416609,175)(6.78400525299968,176)(6.76481425202546,177)(6.74578520096275,178)(6.72691583476742,179)(6.70820393249937,180)(6.6896473162245,181)(6.67124384994991,182)(6.65299143859116,183)(6.63488802697037,184)(6.61693159884427,185)(6.5991201759609,186)(6.58145181714418,187)(6.56392461740526,188)(6.54653670707977,189)(6.52928625099011,190)(6.51217144763179,191)(6.49519052838329,192)(6.47834175673825,193)(6.46162342755964,194)(6.4450338663549,195)(6.42857142857143,196)(6.41223449891187,197)(6.39602149066831,198)(6.37993084507502,199)(6.36396103067893,200)(6.34811054272738,201)(6.33237790257263,202)(6.31676165709237,203)(6.30126037812604,204)(6.2858726619262,205)(6.27059712862456,206)(6.25543242171224,207)(6.24037720753383,208)(6.22543017479467,209)(6.21059003408119,210)(6.19585551739363,211)(6.18122537769101,212)(6.16669838844779,213)(6.15227334322197,214)(6.13794905523426,215)(6.12372435695795,216)(6.10959809971918,217)(6.09556915330737,218)(6.08163640559537,219)(6.06779876216918,220)(6.05405514596681,221)(6.04040449692622,222)(6.02684577164183,223)(6.01337794302955,224)(6,225)(5.98671094713965,226)(5.97350980439975,227)(5.9603956067927,228)(5.94736740409581,229)(5.93442426056208,230)(5.92156525463792,231)(5.90878947868752,232)(5.89609603872377,233)(5.88348405414552,234)(5.87095265748098,235)(5.85850099413707,236)(5.84612822215468,237)(5.83383351196948,238)(5.82161604617836,239)(5.80947501931113,240)(5.79740963760748,241)(5.78541911879903,242)(5.77350269189626,243)(5.76165959698032,244)(5.74988908499946,245)(5.73819041757004,246)(5.726562866782,247)(5.71500571500857,248)(5.7035182547203,249)(5.69209978830308,250)(5.68074962788023,251)(5.66946709513841,252)(5.65825152115738,253)(5.64710224624343,254)(5.63601861976635,255)(5.625,256)(5.61404575396625,257)(5.60315525728221,258)(5.5923278940108,259)(5.58156305651438,260)(5.57086014531156,261)(5.56021856893694,262)(5.54963774380387,263)(5.53911709406997,264)
				
				};

				\addplot [
				color=blue, 
				]
				coordinates {
				(63.6396103067893,1)(45,2)(36.7423461417477,3)(31.8198051533946,4)(28.4604989415154,5)(25.9807621135332,6)(24.0535117721182,7)(22.5,8)(21.2132034355964,9)(20.1246117974981,10)(19.1880644720049,11)(18.3711730708738,12)(17.6504521624366,13)(17.0084012854152,14)(16.431676725155,15)(15.9099025766973,16)(15.4348726628258,17)(15,18)(14.5999279017686,19)(14.2302494707577,20)(13.8873014965883,21)(13.5680105059994,22)
				
				};

			\end{axis}

		\end{tikzpicture}
		
		\caption{$q=0.2, K=100, {_S}C=100, {_H}C=150, r=3\%, \alpha=90, I=2$}
		
	\end{subfigure}

	
	\begin{subfigure}{.5\textwidth}
		\begin{tikzpicture}
			
			\begin{axis}[
				width=6cm,
				axis lines = left,
				xlabel = Premium ($P$),
				ylabel = Numer of customers ($n$),
				xmin=26, xmax=40,
				xtick={26,45},
				ymin=0, ymax=10,
				extra x ticks={31,33.5,40},
				extra x tick labels={${_H}P_U$, ${_S}P_L$, $P_M$},
				ytick={0, 5,10,20,40,60,80,100,120},
				yticklabel style={
					/pgf/number format/fixed,
					/pgf/number format/precision=2
				},
				scaled y ticks=false,
				legend style ={ at={(1.03,1)}, 
					anchor=north west, 
					fill=white,align=left},
				]
				
				\addplot [
			color=red, 
			]
			coordinates {
				
				(-26.966827858044,1)(17.855454708738,2)(29.4862296716193,3)(34.016586070978,4)(36.0778353693702,5)(37.0631163880225,6)(37.514307182532,7)(37.677727354369,8)(37.6777240473187,9)(37.581949852079,10)(37.4293066322697,11)(37.2431148358097,12)(37.0377988639639,13)(36.8224878305501,14)(36.6030506541812,15)(36.383293035489,16)(36.1656853931607,17)(35.951818236246,18)(35.7426939578535,19)(35.5389176846851,20)(35.3408243062823,21)(35.1485642908741,22)(34.9621623871869,23)(34.7815581940112,24)(34.6066344283912,25)(34.4372367422769,26)(34.2731876683176,27)(34.1142964484088,28)(33.9603659512515,29)(33.8111975178207,30)(33.6665943239133,31)(33.5263636771845,32)(33.3903185466971,33)(33.2582785392077,34)(33.1300704770757,35)(33.005528690326,36)(32.8844951049421,37)(32.766819187426,38)(32.6523577896193,39)(32.5409749260395,40)(32.4325415073597,41)(32.3269350472961,42)(32.2240393554582,43)(32.1237442252258,44)(32.0259451231234,45)(31.9305428842369,46)(31.8374434167854,47)(31.7465574179048,48)(31.6578001019121,49)(31.5710909417476,50)(31.4863534238728,51)(31.4035148165973,52)(31.3225059515961,53)(31.2432610182297,54)(31.1657173701789,55)(31.0898153438465,56)(31.0154980879389,57)(30.9427114036303,58)(30.8714035947051,59)(30.8015253270906,60)(30.7330294972024,61)(30.6658711085512,62)(30.6000071560821,63)(30.5353965177445,64)(30.4719998528194,65)(30.4097795065563,66)(30.3486994207011,67)(30.2887250495215,68)(30.229823280964,69)(30.171962362597,70)(30.1151118320223,71)(30.0592424514563,72)(30.004326146203,73)(29.9503359467608,74)(29.8972459343239,75)(29.8450311894531,76)(29.7936677437105,77)(29.7431325340634,78)(29.6934033598791,79)(29.6444588423426,80)(29.5962783861433,81)(29.5488421432856,82)(29.5021309788895,83)(29.4561264388554,84)(29.4108107192785,85)(29.3661666375027,86)(29.3221776047146,87)(29.2788275999825,88)(29.2361011456534,89)(29.1939832840263,90)(29.1524595552255,91)(29.1115159762021,92)(29.0711390207988,93)(29.0313156008142,94)(28.99203304801,95)(28.9532790970056,96)(28.9150418690114,97)(28.8773098563503,98)(28.8400719077263,99)(28.8033172141956,100)(28.7670352958039,101)(28.7312159888509,102)(28.6958494337494,103)(28.6609260634461,104)(28.6264365923739,105)(28.5923720059078,106)(28.5587235502967,107)(28.5254827230477,108)(28.4926412637373,109)(28.4601911452293,110)(28.4281245652768,111)(28.3964339384901,112)(28.3651118886512,113)(28.3341512413568,114)(28.3035450169756,115)(28.2732864239016,116)(28.2433688520905,117)(28.2137858668654,118)(28.1845312029773,119)(28.1555987589104,120)(28.1269825914175,121)(28.0986769102779,122)(28.0706760732645,123)(28.0429745813115,124)(28.015567073874,125)(27.9884483244691,126)(27.9616132363908,127)(27.9350568385922,128)(27.9087742817245,129)(27.8827608343283,130)(27.8570118791693,131)(27.8315229097122,132)(27.8062895267263,133)(27.7813074350179,134)(27.7565724402826,135)(27.7320804460744,136)(27.707827450884,137)(27.6838095453239,138)(27.6600229094143,139)(27.6364638099664,140)(27.6131285980585,141)(27.5900137066013,142)(27.5671156479885,143)(27.5444310118297,144)(27.521956462761,145)(27.4996887383325,146)(27.477624646966,147)(27.4557610659845,148)(27.4340949397064,149)(27.4126232776045,150)(27.3913431525269,151)(27.3702516989761,152)(27.3493461114457,153)(27.3286236428108,154)(27.3080816027719,155)(27.2877173563481,156)(27.2675283224201,157)(27.2475119723189,158)(27.2276658284598,159)(27.2079874630198,160)(27.1884744966565,161)(27.1691245972672,162)(27.1499354787865,163)(27.1309049000213,164)(27.1120306635216,165)(27.0933106144857,166)(27.0747426396991,167)(27.0563246665052,168)(27.0380546618073,169)(27.0199306311,170)(27.0019506175301,171)(26.9841127009849,172)(26.9664149972073,173)(26.9488556569373,174)(26.9314328650778,175)(26.9141448398856,176)(26.8969898321847,177)(26.8799661246027,178)(26.8630720308293,179)(26.846305894895,180)(26.8296660904713,181)(26.8131510201892,182)(26.7967591149783,183)(26.7804888334228,184)(26.7643386611366,185)(26.7483071101549,186)(26.7323927183425,187)(26.7165940488182,188)(26.7009096893957,189)(26.685338252038,190)(26.6698783723281,191)(26.6545287089524,192)(26.6392879431993,193)(26.6241547784697,194)(26.6091279398012,195)(26.594206173405,196)(26.5793882462139,197)(26.5646729454429,198)(26.550059078161,199)(26.5355454708738,200)(26.5211309691169,201)(26.5068144370596,202)(26.4925947571191,203)(26.4784708295835,204)(26.4644415722452,205)(26.4505059200434,206)(26.4366628247145,207)(26.4229112544525,208)(26.4092501935771,209)(26.3956786422099,210)(26.3821956159586,211)(26.3688001456094,212)(26.3554912768258,213)(26.3422680698558,214)(26.3291295992454,215)(26.3160749535593,216)(26.3031032351079,217)(26.2902135596814,218)(26.2774050562895,219)(26.2646768669076,220)(26.252028146229,221)(26.2394580614226,222)(26.2269657918963,223)(26.2145505290661,224)(26.2022114761304,225)(26.1899478478493,226)(26.1777588703291,227)(26.1656437808116,228)(26.1536018274684,229)(26.1416322691998,230)(26.1297343754375,231)(26.1179074259531,232)(26.1061507106693,233)(26.0944635294769,234)(26.0828451920542,235)(26.0712950176915,236)(26.0598123351195,237)(26.0483964823404,238)(26.0370468064637,239)(26.0257626635453,240)(26.0145434184297,241)(26.003388444596,242)(25.9922971240071,243)(25.9812688469619,244)(25.9703030119509,245)(25.9593990255148,246)(25.948556302106,247)(25.937774263953,248)(25.9270523409278,249)(25.9163899704158,250)(25.9057865971885,251)(25.8952416732791,252)(25.8847546578603,253)(25.8743250171246,254)(25.8639522241674,255)(25.8536357588723,256)(25.8433751077983,257)(25.8331697640702,258)(25.8230192272703,259)(25.8129230033328,260)(25.8028806044401,261)(25.7928915489214,262)(25.7829553611528,263)(25.77307157146,264)
				
			};

			\addplot [
			color=black!30!red, name path=VAR
			]
			coordinates {
			(-76.966827858044,1)(-7.14454529126203,2)(12.8195630049527,3)(21.516586070978,4)(26.0778353693702,5)(28.7297830546891,6)(30.3714500396748,7)(31.427727354369,8)(32.1221684917631,9)(32.581949852079,10)(32.8838520868152,11)(33.076448169143,12)(33.19164501781,13)(33.2510592591215,14)(33.2697173208478,15)(33.258293035489,16)(33.2245089225724,17)(33.1740404584682,18)(33.1111150104851,19)(33.0389176846851,20)(32.9598719253299,21)(32.8758370181469,22)(32.7882493437086,23)(32.6982248606779,24)(32.6066344283912,25)(32.5141598191999,26)(32.4213358164657,27)(32.3285821626946,28)(32.236228020217,29)(32.1445308511541,30)(32.0536910981069,31)(31.9638636771845,32)(31.8751670315455,33)(31.7876903039135,34)(31.7014990485042,35)(31.6166398014371,36)(31.5331437535907,37)(31.4510297137418,38)(31.370306507568,39)(31.2909749260395,40)(31.2130293122378,41)(31.1364588568199,42)(31.0612486577838,43)(30.9873805888621,44)(30.9148340120123,45)(30.8435863624977,46)(30.7736136295513,47)(30.7048907512382,48)(30.6373919386468,49)(30.5710909417476,50)(30.5059612670101,51)(30.4419763550589,52)(30.3791097251811,53)(30.3173350923038,54)(30.256626461088,55)(30.1969582009893,56)(30.1383051054828,57)(30.080642438113,58)(30.0239459675864,59)(29.9681919937573,60)(29.9133573660549,61)(29.859419495648,62)(29.8063563624313,63)(29.7541465177445,64)(29.7027690835887,65)(29.6522037489806,66)(29.6024307639846,67)(29.5534309318745,68)(29.5051855998046,69)(29.4576766483113,70)(29.4108864799096,71)(29.3647980070119,72)(29.3193946393537,73)(29.2746602710851,74)(29.2305792676572,75)(29.187136452611,76)(29.1443170943598,77)(29.1021068930378,78)(29.060491967474,79)(29.0194588423426,80)(28.978994435526,81)(28.9390860457247,82)(28.8997213403353,83)(28.8608883436174,84)(28.8225754251609,85)(28.7847712886655,86)(28.7474649610365,87)(28.7106457818007,88)(28.6743033928444,89)(28.6384277284708,90)(28.603009005775,91)(28.5680377153326,92)(28.5335046121967,93)(28.4994007071972,94)(28.4657172585363,95)(28.4324457636723,96)(28.3995779514856,97)(28.3671057747177,98)(28.3350214026758,99)(28.3033172141956,100)(28.2719857908534,101)(28.2410199104196,102)(28.2104125405456,103)(28.1801568326769,104)(28.1502461161835,105)(28.1206738927002,106)(28.0914338306705,107)(28.0625197600847,108)(28.033925667407,109)(28.0056456906838,110)(27.9776741148263,111)(27.9500053670616,112)(27.922634012545,113)(27.8955547501287,114)(27.86876240828,115)(27.842251941143,116)(27.8160184247401,117)(27.790057053306,118)(27.7643631357504,119)(27.7389320922437,120)(27.7137594509216,121)(27.6888408447041,122)(27.6641720082238,123)(27.6397487748599,124)(27.615567073874,125)(27.5916229276437,126)(27.5679124489893,127)(27.5444318385922,128)(27.5211773824997,129)(27.4981454497129,130)(27.4753324898563,131)(27.4527350309243,132)(27.4303496771022,133)(27.4081731066596,134)(27.3862020699122,135)(27.3644333872509,136)(27.3428639472344,137)(27.3214907047442,138)(27.3003106791985,139)(27.2793209528236,140)(27.2585186689805,141)(27.237901030545,142)(27.2174652983382,143)(27.1972087896074,144)(27.1771288765541,145)(27.1572229849078,146)(27.1374885925443,147)(27.1179232281467,148)(27.0985244699077,149)(27.0792899442712,150)(27.0602173247123,151)(27.0413043305551,152)(27.0225487258248,153)(27.0039483181355,154)(26.9855009576106,155)(26.9672045358353,156)(26.9490569848405,157)(26.9310562761164,158)(26.9132004196547,159)(26.8954874630198,160)(26.8779154904453,161)(26.8604826219585,162)(26.8431870125288,163)(26.8260268512408,164)(26.8090003604913,165)(26.7921057952086,166)(26.7753414420943,167)(26.7587056188862,168)(26.7421966736416,169)(26.7258129840412,170)(26.7095529567114,171)(26.6934150265663,172)(26.6773976561669,173)(26.6614993350982,174)(26.6457185793635,175)(26.6300539307947,176)(26.6145039564784,177)(26.5990672481982,178)(26.5837424218907,179)(26.5685281171173,180)(26.5534229965486,181)(26.538425745464,182)(26.5235350712624,183)(26.508749702988,184)(26.4940683908664,185)(26.4794899058539,186)(26.4650130391981,187)(26.4506366020097,188)(26.4363594248454,189)(26.4221803573012,190)(26.408098267616,191)(26.3941120422858,192)(26.3802205856863,193)(26.3664228197068,194)(26.352717683391,195)(26.3391041325887,196)(26.3255811396149,197)(26.3121476929176,198)(26.298802796754,199)(26.2855454708738,200)(26.2723747502114,201)(26.2592896845844,202)(26.2462893383999,203)(26.2333727903678,204)(26.2205391332209,205)(26.2077874734414,206)(26.1951169309947,207)(26.1825266390679,208)(26.1700157438163,209)(26.1575834041146,210)(26.1452287913141,211)(26.1329510890056,212)(26.1207494927882,213)(26.1086232100427,214)(26.0965714597106,215)(26.0845934720778,216)(26.0726884885641,217)(26.0608557615163,218)(26.0490945540064,219)(26.0374041396349,220)(26.0257838023376,221)(26.0142328361974,222)(26.0027505452595,223)(25.9913362433518,224)(25.9799892539082,225)(25.9687089097962,226)(25.9574945531484,227)(25.9463455351975,228)(25.9352612161147,229)(25.924240964852,230)(25.9132841589873,231)(25.9023901845737,232)(25.8915584359912,233)(25.8807883158017,234)(25.8700792346073,235)(25.8594306109119,236)(25.8488418709845,237)(25.8383124487269,238)(25.8278417855432,239)(25.817429330212,240)(25.8070745387616,241)(25.7967768743481,242)(25.7865358071347,243)(25.776350814175,244)(25.7662213792978,245)(25.7561469929944,246)(25.7461271523084,247)(25.7361613607272,248)(25.7262491280764,249)(25.7163899704158,250)(25.7065834099375,251)(25.6968289748664,252)(25.6871261993623,253)(25.6774746234238,254)(25.6678737927949,255)(25.6583232588723,256)(25.6488225786154,257)(25.6393713144578,258)(25.6299690342201,259)(25.6206153110251,260)(25.6113097232141,261)(25.6020518542649,262)(25.5928412927117,263)(25.583677632066,264)

			};
			
			\addplot [
			color=blue, 
			]
			coordinates {
				(90,1)(63.6396103067893,2)(51.9615242270663,3)(45,4)(40.2492235949962,5)(36.7423461417477,6)(34.0168025708304,7)(31.8198051533946,8)(30,9)(28.4604989415154,10)(27.1360210119987,11)(25.9807621135332,12)(24.9615088301353,13)(24.0535117721182,14)(23.2379000772445,15)(22.5,16)(21.82820625327,17)(21.2132034355964,18)(20.6474160483506,19)(20.1246117974981,20)(19.6396101212393,21)(19.1880644720049,22)(18.7662972651367,23)(18.3711730708738,24)(18,25)(17.6504521624366,26)(17.3205080756888,27)(17.0084012854152,28)(16.7125804359347,29)(16.431676725155,30)(16.1644771824097,31)(15.9099025766973,32)(15.6669890360128,33)(15.4348726628258,34)(15.2127765851133,35)(15,36)(14.7959088574822,37)(14.5999279017686,38)(14.4115338424578,39)(14.2302494707577,40)(14.0556385699745,41)(13.8873014965883,42)(13.7248713299344,43)(13.5680105059994,44)(13.4164078649987,45)(13.2697760539407,46)(13.1278492348105,47)(12.9903810567666,48)(12.8571428571429,49)(12.7279220613579,50)(12.6025207562521,51)(12.4807544150677,52)(12.362450755382,53)(12.2474487139159,54)(12.1355975243384,55)(12.0267558860591,56)(11.9207912135854,57)(11.817578957375,58)(11.7170019882741,59)(11.6189500386223,60)(11.5233191939606,61)(11.4300114300171,62)(11.3389341902768,63)(11.25,64)(11.1631261130288,65)(11.0782341881399,66)(10.9952499920675,67)(10.914103126635,68)(10.8347267777192,69)(10.7570574840095,70)(10.6810349237447,71)(10.6066017177982,72)(10.5337032476518,73)(10.4622874869437,74)(10.3923048454133,75)(10.3237080241753,76)(10.2564518813674,77)(10.1904933073014,78)(10.1257911083342,79)(10.0623058987491,80)(10,81)(9.93883734673619,82)(9.87878339907213,83)(9.81980506061966,84)(9.76187060183953,85)(9.70494958830946,86)(9.64901281354015,87)(9.59403223600247,88)(9.53998092005724,89)(9.48683298050514,90)(9.43456353049726,91)(9.38314863256837,92)(9.33256525257383,93)(9.28279121632914,94)(9.23380516876639,95)(9.18558653543692,96)(9.13811548620257,97)(9.0913729009699,98)(9.04534033733291,99)(9,100)(8.9553347118899,101)(8.91132788679007,102)(8.86796350347864,103)(8.82522608121828,104)(8.7831006565368,105)(8.74157276121538,106)(8.70062840141097,107)(8.66025403784439,108)(8.62043656699036,109)(8.58116330321033,110)(8.54242196177249,111)(8.50420064270761,112)(8.46648781545237,113)(8.42927230423525,114)(8.39254327416282,115)(8.35629021796733,116)(8.32050294337844,117)(8.28517156108491,118)(8.2502864732539,119)(8.21583836257749,120)(8.18181818181818,121)(8.14821714382667,122)(8.11502671200689,123)(8.08223859120487,124)(8.04984471899924,125)(8.01783725737273,126)(7.98620858474502,127)(7.95495128834866,128)(7.92405815693061,129)(7.89352217376326,130)(7.86333650994934,131)(7.8334945180064,132)(7.80398972571708,133)(7.77481583023224,134)(7.74596669241483,135)(7.7174363314129,136)(7.68921891945085,137)(7.66130877682874,138)(7.63370036711974,139)(7.60638829255665,140)(7.57936728959867,141)(7.5526322246702,142)(7.52617809006382,143)(7.5,144)(7.4740931868366,145)(7.44845299742131,146)(7.4230748895809,147)(7.39795442874108,148)(7.37308728467136,149)(7.34846922834953,150)(7.32409612894044,151)(7.29996395088431,152)(7.27606875108999,153)(7.25240667622842,154)(7.22897396012249,155)(7.20576692122892,156)(7.1827819602086,157)(7.16001555758157,158)(7.1374642714633,159)(7.11512473537885,160)(7.09299365615191,161)(7.07106781186548,162)(7.04934404989162,163)(7.02781928498727,164)(7.00649049745371,165)(6.985354731357,166)(6.96440909280723,167)(6.94365074829414,168)(6.92307692307692,169)(6.90268489962633,170)(6.88247201611685,171)(6.86243566496721,172)(6.84257329142735,173)(6.82288239221013,174)(6.80336051416609,175)(6.78400525299968,176)(6.76481425202546,177)(6.74578520096275,178)(6.72691583476742,179)(6.70820393249937,180)(6.6896473162245,181)(6.67124384994991,182)(6.65299143859116,183)(6.63488802697037,184)(6.61693159884427,185)(6.5991201759609,186)(6.58145181714418,187)(6.56392461740526,188)(6.54653670707977,189)(6.52928625099011,190)(6.51217144763179,191)(6.49519052838329,192)(6.47834175673825,193)(6.46162342755964,194)(6.4450338663549,195)(6.42857142857143,196)(6.41223449891187,197)(6.39602149066831,198)(6.37993084507502,199)(6.36396103067893,200)(6.34811054272738,201)(6.33237790257263,202)(6.31676165709237,203)(6.30126037812604,204)(6.2858726619262,205)(6.27059712862456,206)(6.25543242171224,207)(6.24037720753383,208)(6.22543017479467,209)(6.21059003408119,210)(6.19585551739363,211)(6.18122537769101,212)(6.16669838844779,213)(6.15227334322197,214)(6.13794905523426,215)(6.12372435695795,216)(6.10959809971918,217)(6.09556915330737,218)(6.08163640559537,219)(6.06779876216918,220)(6.05405514596681,221)(6.04040449692622,222)(6.02684577164183,223)(6.01337794302955,224)(6,225)(5.98671094713965,226)(5.97350980439975,227)(5.9603956067927,228)(5.94736740409581,229)(5.93442426056208,230)(5.92156525463792,231)(5.90878947868752,232)(5.89609603872377,233)(5.88348405414552,234)(5.87095265748098,235)(5.85850099413707,236)(5.84612822215468,237)(5.83383351196948,238)(5.82161604617836,239)(5.80947501931113,240)(5.79740963760748,241)(5.78541911879903,242)(5.77350269189626,243)(5.76165959698032,244)(5.74988908499946,245)(5.73819041757004,246)(5.726562866782,247)(5.71500571500857,248)(5.7035182547203,249)(5.69209978830308,250)(5.68074962788023,251)(5.66946709513841,252)(5.65825152115738,253)(5.64710224624343,254)(5.63601861976635,255)(5.625,256)(5.61404575396625,257)(5.60315525728221,258)(5.5923278940108,259)(5.58156305651438,260)(5.57086014531156,261)(5.56021856893694,262)(5.54963774380387,263)(5.53911709406997,264)
				
			};

			\addplot [
			color=blue, 
			]
			coordinates {
				(63.6396103067893,1)(45,2)(36.7423461417477,3)(31.8198051533946,4)(28.4604989415154,5)(25.9807621135332,6)(24.0535117721182,7)(22.5,8)(21.2132034355964,9)(20.1246117974981,10)(19.1880644720049,11)(18.3711730708738,12)(17.6504521624366,13)(17.0084012854152,14)(16.431676725155,15)(15.9099025766973,16)(15.4348726628258,17)(15,18)(14.5999279017686,19)(14.2302494707577,20)(13.8873014965883,21)(13.5680105059994,22)
				
			};

		\end{axis}

\end{tikzpicture}

\caption{$q=0.2, K=100, {_S}C=150, {_H}C=200, r=3\%, \alpha=90, I=2$}
		
	\end{subfigure}
	
\begin{subfigure}{.5\textwidth}
	\begin{tikzpicture}
		
		\begin{axis}[
			width=6cm,
			axis lines = left,
			xlabel = Premium ($P$),
			ylabel = Numer of customers ($n$),
			xmin=26, xmax=40,
			xtick={26,45},
			ymin=0, ymax=10,
			extra x ticks={33,36,40},
			extra x tick labels={ ${_S}P_L$,${_H}P_U$, $P_M$},
			ytick={0, 5,10,20,40,60,80,100,120},
			yticklabel style={
				/pgf/number format/fixed,
				/pgf/number format/precision=2
			},
			scaled y ticks=false,
			legend style ={ at={(1.03,1)}, 
				anchor=north west, 
				fill=white,align=left},
			]
			
			\addplot [
			color=red, 
			]
			coordinates {
				
				(-26.966827858044,1)(17.855454708738,2)(29.4862296716193,3)(34.016586070978,4)(36.0778353693702,5)(37.0631163880225,6)(37.514307182532,7)(37.677727354369,8)(37.6777240473187,9)(37.581949852079,10)(37.4293066322697,11)(37.2431148358097,12)(37.0377988639639,13)(36.8224878305501,14)(36.6030506541812,15)(36.383293035489,16)(36.1656853931607,17)(35.951818236246,18)(35.7426939578535,19)(35.5389176846851,20)(35.3408243062823,21)(35.1485642908741,22)(34.9621623871869,23)(34.7815581940112,24)(34.6066344283912,25)(34.4372367422769,26)(34.2731876683176,27)(34.1142964484088,28)(33.9603659512515,29)(33.8111975178207,30)(33.6665943239133,31)(33.5263636771845,32)(33.3903185466971,33)(33.2582785392077,34)(33.1300704770757,35)(33.005528690326,36)(32.8844951049421,37)(32.766819187426,38)(32.6523577896193,39)(32.5409749260395,40)(32.4325415073597,41)(32.3269350472961,42)(32.2240393554582,43)(32.1237442252258,44)(32.0259451231234,45)(31.9305428842369,46)(31.8374434167854,47)(31.7465574179048,48)(31.6578001019121,49)(31.5710909417476,50)(31.4863534238728,51)(31.4035148165973,52)(31.3225059515961,53)(31.2432610182297,54)(31.1657173701789,55)(31.0898153438465,56)(31.0154980879389,57)(30.9427114036303,58)(30.8714035947051,59)(30.8015253270906,60)(30.7330294972024,61)(30.6658711085512,62)(30.6000071560821,63)(30.5353965177445,64)(30.4719998528194,65)(30.4097795065563,66)(30.3486994207011,67)(30.2887250495215,68)(30.229823280964,69)(30.171962362597,70)(30.1151118320223,71)(30.0592424514563,72)(30.004326146203,73)(29.9503359467608,74)(29.8972459343239,75)(29.8450311894531,76)(29.7936677437105,77)(29.7431325340634,78)(29.6934033598791,79)(29.6444588423426,80)(29.5962783861433,81)(29.5488421432856,82)(29.5021309788895,83)(29.4561264388554,84)(29.4108107192785,85)(29.3661666375027,86)(29.3221776047146,87)(29.2788275999825,88)(29.2361011456534,89)(29.1939832840263,90)(29.1524595552255,91)(29.1115159762021,92)(29.0711390207988,93)(29.0313156008142,94)(28.99203304801,95)(28.9532790970056,96)(28.9150418690114,97)(28.8773098563503,98)(28.8400719077263,99)(28.8033172141956,100)(28.7670352958039,101)(28.7312159888509,102)(28.6958494337494,103)(28.6609260634461,104)(28.6264365923739,105)(28.5923720059078,106)(28.5587235502967,107)(28.5254827230477,108)(28.4926412637373,109)(28.4601911452293,110)(28.4281245652768,111)(28.3964339384901,112)(28.3651118886512,113)(28.3341512413568,114)(28.3035450169756,115)(28.2732864239016,116)(28.2433688520905,117)(28.2137858668654,118)(28.1845312029773,119)(28.1555987589104,120)(28.1269825914175,121)(28.0986769102779,122)(28.0706760732645,123)(28.0429745813115,124)(28.015567073874,125)(27.9884483244691,126)(27.9616132363908,127)(27.9350568385922,128)(27.9087742817245,129)(27.8827608343283,130)(27.8570118791693,131)(27.8315229097122,132)(27.8062895267263,133)(27.7813074350179,134)(27.7565724402826,135)(27.7320804460744,136)(27.707827450884,137)(27.6838095453239,138)(27.6600229094143,139)(27.6364638099664,140)(27.6131285980585,141)(27.5900137066013,142)(27.5671156479885,143)(27.5444310118297,144)(27.521956462761,145)(27.4996887383325,146)(27.477624646966,147)(27.4557610659845,148)(27.4340949397064,149)(27.4126232776045,150)(27.3913431525269,151)(27.3702516989761,152)(27.3493461114457,153)(27.3286236428108,154)(27.3080816027719,155)(27.2877173563481,156)(27.2675283224201,157)(27.2475119723189,158)(27.2276658284598,159)(27.2079874630198,160)(27.1884744966565,161)(27.1691245972672,162)(27.1499354787865,163)(27.1309049000213,164)(27.1120306635216,165)(27.0933106144857,166)(27.0747426396991,167)(27.0563246665052,168)(27.0380546618073,169)(27.0199306311,170)(27.0019506175301,171)(26.9841127009849,172)(26.9664149972073,173)(26.9488556569373,174)(26.9314328650778,175)(26.9141448398856,176)(26.8969898321847,177)(26.8799661246027,178)(26.8630720308293,179)(26.846305894895,180)(26.8296660904713,181)(26.8131510201892,182)(26.7967591149783,183)(26.7804888334228,184)(26.7643386611366,185)(26.7483071101549,186)(26.7323927183425,187)(26.7165940488182,188)(26.7009096893957,189)(26.685338252038,190)(26.6698783723281,191)(26.6545287089524,192)(26.6392879431993,193)(26.6241547784697,194)(26.6091279398012,195)(26.594206173405,196)(26.5793882462139,197)(26.5646729454429,198)(26.550059078161,199)(26.5355454708738,200)(26.5211309691169,201)(26.5068144370596,202)(26.4925947571191,203)(26.4784708295835,204)(26.4644415722452,205)(26.4505059200434,206)(26.4366628247145,207)(26.4229112544525,208)(26.4092501935771,209)(26.3956786422099,210)(26.3821956159586,211)(26.3688001456094,212)(26.3554912768258,213)(26.3422680698558,214)(26.3291295992454,215)(26.3160749535593,216)(26.3031032351079,217)(26.2902135596814,218)(26.2774050562895,219)(26.2646768669076,220)(26.252028146229,221)(26.2394580614226,222)(26.2269657918963,223)(26.2145505290661,224)(26.2022114761304,225)(26.1899478478493,226)(26.1777588703291,227)(26.1656437808116,228)(26.1536018274684,229)(26.1416322691998,230)(26.1297343754375,231)(26.1179074259531,232)(26.1061507106693,233)(26.0944635294769,234)(26.0828451920542,235)(26.0712950176915,236)(26.0598123351195,237)(26.0483964823404,238)(26.0370468064637,239)(26.0257626635453,240)(26.0145434184297,241)(26.003388444596,242)(25.9922971240071,243)(25.9812688469619,244)(25.9703030119509,245)(25.9593990255148,246)(25.948556302106,247)(25.937774263953,248)(25.9270523409278,249)(25.9163899704158,250)(25.9057865971885,251)(25.8952416732791,252)(25.8847546578603,253)(25.8743250171246,254)(25.8639522241674,255)(25.8536357588723,256)(25.8433751077983,257)(25.8331697640702,258)(25.8230192272703,259)(25.8129230033328,260)(25.8028806044401,261)(25.7928915489214,262)(25.7829553611528,263)(25.77307157146,264)
				
			};

			\addplot [
			color=black!30!red, name path=VAR
			]
			coordinates {
			(-36.966827858044,1)(12.855454708738,2)(26.152896338286,3)(31.516586070978,4)(34.0778353693702,5)(35.3964497213558,6)(36.0857357539605,7)(36.427727354369,8)(36.5666129362076,9)(36.581949852079,10)(36.5202157231788,11)(36.4097815024763,12)(36.2685680947331,13)(36.1082021162643,14)(35.9363839875145,15)(35.758293035489,16)(35.577450099043,17)(35.3962626806904,18)(35.2163781683798,19)(35.0389176846851,20)(34.8646338300918,21)(34.6940188363287,22)(34.5273797784912,23)(34.3648915273446,24)(34.2066344283912,25)(34.0526213576615,26)(33.9028172979472,27)(33.757153591266,28)(33.6155383650446,29)(33.4778641844874,30)(33.3440136787521,31)(33.2138636771845,32)(33.0872882436668,33)(32.9641608921488,34)(32.8443561913614,35)(32.7277509125482,36)(32.6142248346718,37)(32.5036612926891,38)(32.395947533209,39)(32.2909749260395,40)(32.1886390683353,41)(32.0888398092009,42)(31.9914812159233,43)(31.896471497953,44)(31.8037229009012,45)(31.713151579889,46)(31.6246774593386,47)(31.5382240845715,48)(31.453718469259,49)(31.3710909417476,50)(31.2902749925003,51)(31.2112071242896,52)(31.1338267063131,53)(31.0580758330445,54)(30.9838991883607,55)(30.911243915275,56)(30.8400594914477,57)(30.7702976105268,58)(30.7019120692813,59)(30.6348586604239,60)(30.5690950709729,61)(30.5045807859706,62)(30.4412769973519,63)(30.3791465177445,64)(30.3181536989733,65)(30.2582643550412,66)(30.1994456893578,67)(30.1416662259921,68)(30.0848957447321,69)(30.0291052197398,70)(29.9742667615998,71)(29.9203535625674,72)(29.8673398448331,73)(29.8152008116257,74)(29.7639126009905,75)(29.7134522420846,76)(29.6637976138403,77)(29.6149274058583,78)(29.5668210813981,79)(29.5194588423426,80)(29.4728215960198,81)(29.4268909237734,82)(29.3816490511787,83)(29.3370788198078,84)(29.293163660455,85)(29.2498875677353,86)(29.207235075979,87)(29.1651912363462,88)(29.1237415950916,89)(29.0828721729152,90)(29.0425694453354,91)(29.0028203240282,92)(28.9636121390784,93)(28.9249326220908,94)(28.8867698901152,95)(28.8491124303389,96)(28.8119490855062,97)(28.7752690400238,98)(28.7390618067162,99)(28.7033172141956,100)(28.6680253948138,101)(28.6331767731647,102)(28.5987620551087,103)(28.5647722172923,104)(28.5311984971358,105)(28.4980323832663,106)(28.4652656063715,107)(28.4328901304551,108)(28.4008981444712,109)(28.3692820543202,110)(28.3380344751867,111)(28.3071482242044,112)(28.2766163134299,113)(28.2464319431112,114)(28.2165884952365,115)(28.1870795273499,116)(28.1578987666204,117)(28.1290401041535,118)(28.100497589532,119)(28.072265425577,120)(28.0443379633183,121)(28.0167096971632,122)(27.9893752602563,123)(27.9623294200212,124)(27.935567073874,125)(27.909083245104,126)(27.8828730789105,127)(27.8569318385922,128)(27.8312549018795,129)(27.8058377574052,130)(27.7806760013067,131)(27.7557653339546,132)(27.7311015568015,133)(27.7066805693462,134)(27.6824983662085,135)(27.6585510343097,136)(27.6348347501541,137)(27.611345777208,138)(27.5880804633711,139)(27.5650352385378,140)(27.5422066122429,141)(27.51959117139,142)(27.4971855780585,143)(27.4749865673852,144)(27.4529909455197,145)(27.4311955876475,146)(27.4095974360817,147)(27.388193498417,148)(27.3669808457466,149)(27.3459566109378,150)(27.325117986964,151)(27.3044622252919,152)(27.2839866343215,153)(27.2636885778758,154)(27.2435654737396,155)(27.2236147922455,156)(27.2038340549042,157)(27.1842208330784,158)(27.1647727466988,159)(27.1454874630198,160)(27.1263626954142,161)(27.1073962022055,162)(27.088585785535,163)(27.0699292902652,164)(27.0514246029155,165)(27.0330696506303,166)(27.0148624001781,167)(26.9968008569814,168)(26.9788830641741,169)(26.9611071016882,170)(26.9434710853664,171)(26.9259731661012,172)(26.9086115289992,173)(26.8913843925695,174)(26.874290007935,175)(26.8573266580674,176)(26.8404926570434,177)(26.8237863493218,178)(26.8072061090416,179)(26.7907503393395,180)(26.7744174716868,181)(26.7582059652442,182)(26.7421143062351,183)(26.7261410073358,184)(26.7102846070826,185)(26.6945436692947,186)(26.6789167825136,187)(26.6634025594565,188)(26.6479996364856,189)(26.6327066730907,190)(26.6175223513857,191)(26.6024453756191,192)(26.5874744716967,193)(26.5726083867171,194)(26.5578458885192,195)(26.5431857652418,196)(26.5286268248941,197)(26.5141678949378,198)(26.4998078218796,199)(26.4855454708738,200)(26.4713797253358,201)(26.4573094865646,202)(26.4433336733752,203)(26.4294512217403,204)(26.4156610844404,205)(26.401962230723,206)(26.3883536459705,207)(26.3748343313756,208)(26.361403303625,209)(26.3480595945908,210)(26.3348022510297,211)(26.3216303342886,212)(26.3085429200183,213)(26.2955390978932,214)(26.2826179713385,215)(26.269778657263,216)(26.2570202857992,217)(26.2443420000484,218)(26.2317429558329,219)(26.2192223214531,220)(26.2067792774508,221)(26.1944130163775,222)(26.1821227425689,223)(26.1699076719232,224)(26.157767031686,225)(26.1457000602387,226)(26.1337060068929,227)(26.1217841316888,228)(26.1099337051977,229)(26.0981540083302,230)(26.0864443321475,231)(26.0748039776772,232)(26.0632322557337,233)(26.0517284867419,234)(26.0402920005648,235)(26.0289221363356,236)(26.0176182422925,237)(26.0063796756177,238)(25.9952058022796,239)(25.9840959968786,240)(25.9730496424961,241)(25.9620661305464,242)(25.9511448606326,243)(25.9402852404045,244)(25.9294866854202,245)(25.9187486190107,246)(25.9080704721465,247)(25.8974516833079,248)(25.8868916983576,249)(25.8763899704158,250)(25.8659459597383,251)(25.8555591335966,252)(25.8452289661607,253)(25.8349549383844,254)(25.8247365378929,255)(25.8145732588723,256)(25.8044646019617,257)(25.7944100741477,258)(25.7844091886602,259)(25.7744614648713,260)(25.7645664281949,261)(25.7547236099901,262)(25.7449325474646,263)(25.7351927835812,264)

			};
			
			\addplot [
			color=blue, 
			]
			coordinates {
				(90,1)(63.6396103067893,2)(51.9615242270663,3)(45,4)(40.2492235949962,5)(36.7423461417477,6)(34.0168025708304,7)(31.8198051533946,8)(30,9)(28.4604989415154,10)(27.1360210119987,11)(25.9807621135332,12)(24.9615088301353,13)(24.0535117721182,14)(23.2379000772445,15)(22.5,16)(21.82820625327,17)(21.2132034355964,18)(20.6474160483506,19)(20.1246117974981,20)(19.6396101212393,21)(19.1880644720049,22)(18.7662972651367,23)(18.3711730708738,24)(18,25)(17.6504521624366,26)(17.3205080756888,27)(17.0084012854152,28)(16.7125804359347,29)(16.431676725155,30)(16.1644771824097,31)(15.9099025766973,32)(15.6669890360128,33)(15.4348726628258,34)(15.2127765851133,35)(15,36)(14.7959088574822,37)(14.5999279017686,38)(14.4115338424578,39)(14.2302494707577,40)(14.0556385699745,41)(13.8873014965883,42)(13.7248713299344,43)(13.5680105059994,44)(13.4164078649987,45)(13.2697760539407,46)(13.1278492348105,47)(12.9903810567666,48)(12.8571428571429,49)(12.7279220613579,50)(12.6025207562521,51)(12.4807544150677,52)(12.362450755382,53)(12.2474487139159,54)(12.1355975243384,55)(12.0267558860591,56)(11.9207912135854,57)(11.817578957375,58)(11.7170019882741,59)(11.6189500386223,60)(11.5233191939606,61)(11.4300114300171,62)(11.3389341902768,63)(11.25,64)(11.1631261130288,65)(11.0782341881399,66)(10.9952499920675,67)(10.914103126635,68)(10.8347267777192,69)(10.7570574840095,70)(10.6810349237447,71)(10.6066017177982,72)(10.5337032476518,73)(10.4622874869437,74)(10.3923048454133,75)(10.3237080241753,76)(10.2564518813674,77)(10.1904933073014,78)(10.1257911083342,79)(10.0623058987491,80)(10,81)(9.93883734673619,82)(9.87878339907213,83)(9.81980506061966,84)(9.76187060183953,85)(9.70494958830946,86)(9.64901281354015,87)(9.59403223600247,88)(9.53998092005724,89)(9.48683298050514,90)(9.43456353049726,91)(9.38314863256837,92)(9.33256525257383,93)(9.28279121632914,94)(9.23380516876639,95)(9.18558653543692,96)(9.13811548620257,97)(9.0913729009699,98)(9.04534033733291,99)(9,100)(8.9553347118899,101)(8.91132788679007,102)(8.86796350347864,103)(8.82522608121828,104)(8.7831006565368,105)(8.74157276121538,106)(8.70062840141097,107)(8.66025403784439,108)(8.62043656699036,109)(8.58116330321033,110)(8.54242196177249,111)(8.50420064270761,112)(8.46648781545237,113)(8.42927230423525,114)(8.39254327416282,115)(8.35629021796733,116)(8.32050294337844,117)(8.28517156108491,118)(8.2502864732539,119)(8.21583836257749,120)(8.18181818181818,121)(8.14821714382667,122)(8.11502671200689,123)(8.08223859120487,124)(8.04984471899924,125)(8.01783725737273,126)(7.98620858474502,127)(7.95495128834866,128)(7.92405815693061,129)(7.89352217376326,130)(7.86333650994934,131)(7.8334945180064,132)(7.80398972571708,133)(7.77481583023224,134)(7.74596669241483,135)(7.7174363314129,136)(7.68921891945085,137)(7.66130877682874,138)(7.63370036711974,139)(7.60638829255665,140)(7.57936728959867,141)(7.5526322246702,142)(7.52617809006382,143)(7.5,144)(7.4740931868366,145)(7.44845299742131,146)(7.4230748895809,147)(7.39795442874108,148)(7.37308728467136,149)(7.34846922834953,150)(7.32409612894044,151)(7.29996395088431,152)(7.27606875108999,153)(7.25240667622842,154)(7.22897396012249,155)(7.20576692122892,156)(7.1827819602086,157)(7.16001555758157,158)(7.1374642714633,159)(7.11512473537885,160)(7.09299365615191,161)(7.07106781186548,162)(7.04934404989162,163)(7.02781928498727,164)(7.00649049745371,165)(6.985354731357,166)(6.96440909280723,167)(6.94365074829414,168)(6.92307692307692,169)(6.90268489962633,170)(6.88247201611685,171)(6.86243566496721,172)(6.84257329142735,173)(6.82288239221013,174)(6.80336051416609,175)(6.78400525299968,176)(6.76481425202546,177)(6.74578520096275,178)(6.72691583476742,179)(6.70820393249937,180)(6.6896473162245,181)(6.67124384994991,182)(6.65299143859116,183)(6.63488802697037,184)(6.61693159884427,185)(6.5991201759609,186)(6.58145181714418,187)(6.56392461740526,188)(6.54653670707977,189)(6.52928625099011,190)(6.51217144763179,191)(6.49519052838329,192)(6.47834175673825,193)(6.46162342755964,194)(6.4450338663549,195)(6.42857142857143,196)(6.41223449891187,197)(6.39602149066831,198)(6.37993084507502,199)(6.36396103067893,200)(6.34811054272738,201)(6.33237790257263,202)(6.31676165709237,203)(6.30126037812604,204)(6.2858726619262,205)(6.27059712862456,206)(6.25543242171224,207)(6.24037720753383,208)(6.22543017479467,209)(6.21059003408119,210)(6.19585551739363,211)(6.18122537769101,212)(6.16669838844779,213)(6.15227334322197,214)(6.13794905523426,215)(6.12372435695795,216)(6.10959809971918,217)(6.09556915330737,218)(6.08163640559537,219)(6.06779876216918,220)(6.05405514596681,221)(6.04040449692622,222)(6.02684577164183,223)(6.01337794302955,224)(6,225)(5.98671094713965,226)(5.97350980439975,227)(5.9603956067927,228)(5.94736740409581,229)(5.93442426056208,230)(5.92156525463792,231)(5.90878947868752,232)(5.89609603872377,233)(5.88348405414552,234)(5.87095265748098,235)(5.85850099413707,236)(5.84612822215468,237)(5.83383351196948,238)(5.82161604617836,239)(5.80947501931113,240)(5.79740963760748,241)(5.78541911879903,242)(5.77350269189626,243)(5.76165959698032,244)(5.74988908499946,245)(5.73819041757004,246)(5.726562866782,247)(5.71500571500857,248)(5.7035182547203,249)(5.69209978830308,250)(5.68074962788023,251)(5.66946709513841,252)(5.65825152115738,253)(5.64710224624343,254)(5.63601861976635,255)(5.625,256)(5.61404575396625,257)(5.60315525728221,258)(5.5923278940108,259)(5.58156305651438,260)(5.57086014531156,261)(5.56021856893694,262)(5.54963774380387,263)(5.53911709406997,264)
				
			};

			\addplot [
			color=blue, 
			]
			coordinates {
				(63.6396103067893,1)(45,2)(36.7423461417477,3)(31.8198051533946,4)(28.4604989415154,5)(25.9807621135332,6)(24.0535117721182,7)(22.5,8)(21.2132034355964,9)(20.1246117974981,10)(19.1880644720049,11)(18.3711730708738,12)(17.6504521624366,13)(17.0084012854152,14)(16.431676725155,15)(15.9099025766973,16)(15.4348726628258,17)(15,18)(14.5999279017686,19)(14.2302494707577,20)(13.8873014965883,21)(13.5680105059994,22)
				
			};

		\end{axis}

	\end{tikzpicture}
	
	\caption{$q=0.2, K=100, {_S}C=150, {_H}C=160, r=3\%, \alpha=90, I=2$}
	
\end{subfigure}
	\caption{Illustration of the cases of positive expected profit.}
	\label{fig:Cases}
\end{figure}

\section{Endogenous capital} \label{sec:endog}

\red{In Section \ref{sec:short} we investigated the possible Nash equilibrium assuming fix capital level. In a concrete time, the capital level is really fixed for
the insurance company, but its level set by the owner at foundation of the company, and even after foundation the level of capital can be modified. In this section
we extend our investigation to situations when companies can decide on capital level. In section \ref{sec:expost} we investigate the ex post capital adjustment, i.e. when
companies get such many client that its capital is not sufficient, it can raise its capital level. In section \ref{sec:expost} we investigate a two period game: in the first period
owners found its companies (and decide on the capital level), and in the second period they compete assuming fix capital level.}

\subsection{Ex post capital adjustment} \label{sec:expost}

\red{In this section we investigate the case, where the companies can adjust its capital level afterward. We assume that all of the companies has the
same amount of capital. They set the premium level simultaneously. If one (or more) company gets so many clients that its capital level is insufficient, it
can adjust its capital to the required level. In insurance industry capital adjustment can appear, but we do not consider it a regular process. First, raising
capital is a time consuming process. For insurance companies the possibility of bond issue is limited, share issue is a longer process, and requires many calculation from the company
which is a costly action. Afterward capital change carries the risk of supervisory inventory which could cause a downfall in company's reputation (which cause financial losses to the company). On the other hand allowing ex post capital adjustment would urge players to found its company with 0 capital, and adjust it, if they get some market
share, but this kind of operation is totally against the regulator's will. So although we do not consider this setting as a regular operations in insurance industry,
we feel it is worth investigating this case and we will see this kind of setting push the market to a monopolistic market, which is undesirable; moreover all the companies suffer losses in this market framework.} 

\red{In this section we investigate the case when the capital can be increased but not decreased (we think that all of the companies was founded with minimal possible capital level).}

\red{If ex post capital adjustment is possible we have to modify the profit function. In Section \ref{sec:mod} we assumed, that }\(A\)\red{\ is so big, that company
avoid it at any time. Considering this the rational assumption is this section is that companies always adjust its capital level, if it is insufficient.
We think that cost is valid in our case: i.e. cost of share issue and increased level of calculation, decrease in firm's reputation. Since multitude of
fix cost is differ, we denote it with }\(B\)\red{.}

\[\pi_i(P_i)
\left\{
\begin{array}{ll}
	\green{n_i(P_i, P_{-i})}(P_i-qK) - r C \ , & \mbox{ if }\hspace{1em} MCR(\green{n_i(P_i, P_{-i})},P_i) \le C \\
\begin{array}{l}	\green{n_i(P_i, P_{-i})}(P_i-qK) - r C-B \\
\hspace{3em}
- r(MCR(\green{n_i(P_i, P_{-i})},P_i) - C)\ ,
\end{array}
\hspace{1em} & \mbox{ if }\hspace{1em} MCR(\green{n_i(P_i, P_{-i})},P_i) > C \\
\end{array}
\right.
\]

\red{It is worth mentioning that \(rC\) is considered sunk cost, however the company decide on the level of incease, so }\(r(MCR(\green{n_i(P_i, P_{-i})},P_i) - C)\)\red{\ cannot
be considered as sunk cost.}

\red{To mover further we have to introduce zero-profit curves. Zero-profit curves depend on capital change (}\(\Delta C\)\red{) and gives the pairs of }\((P,n)\)\red{\ which
ensures exactly as much profit as needed to cover variable costs, i.e. }\(B+r\Delta C\)\red{. Zero-profit curve }\((ZP(P,\Delta C))\)\red{\ for fixed
capital change is a hyperbolic function:}
\[
ZP(P,\Delta C)=\frac{B+r\Delta C}{P-qK}
\] 
 
\begin{figure}[ht!]
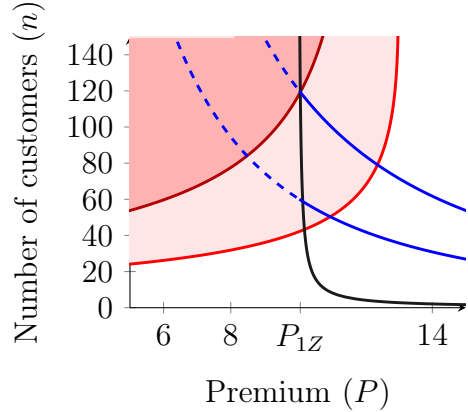

	\centering
	\pgfplotsset{scaled y ticks=false}
	\pgfplotsset{every axis plot/.append style={line width=1.1pt}}

	
	\caption{\footnotesize{$q=0.1, K=100, C^S=500, C^H=835.75, r=1\%, \alpha=110, I=2, B=5$}}
	\label{fig:P1Z}

\end{figure}

\red{It is quite clear, that as increase }\(\Delta C\)\red{\ zero-profit curve moves upward but in case of hyperbolic function this can also be interpreted as 
a rightward move. On the other hand an increase in} \(\Delta C\)\red{\ moves to the MPR curve to the left. So if the point of intersection of zero-profit curve and MPR curve ensures positive profit, with increasing }\(\Delta C\)\red{\ 
we can reach that this point of intersection should result zero profit, in other words: the three curves intersect each other in a common point}\(P_{1Z}\)\red{\ (see Figure \ref{fig:P1Z}). In a formal way we prove it in Proposition \ref{prop:intsectZP}.}

\begin{prop}\label{prop:intsectZP}
\red{The MPR curve and the zero profit curve has unique point of intersection for a fixed }\(\Delta C\)\red{\ value, where the number
of policies is}
\begin{equation} \label{eq:ZPn}
n_{ZMPR}(\Delta C)=\left(\frac{B+C+(1+r)\Delta C}{\phi\sqrt{q(1-q)}K}\right)^2 \ ,
\end{equation}
\red{and the premium is}
\begin{equation} \label{eq:ZPP}
P_{ZMPR}(\Delta C)=qK+\frac{B+r\Delta C}{\left(\frac{B+C+(1+r)\Delta C}{\phi\sqrt{q(1-q)}K}\right)^2}
\end{equation}
\end{prop}

\begin{proof}
\red{Since it is better to work with MPR function, we take the inverse of zero-profit function:}
\[
ZP^{-1}(n,\Delta C)=qK+\frac{B+r\Delta C}{n} \ .
\]

\red{Moving forward we make equal the two function:}
\[
qK+\frac{B+r\Delta C}{n}=qK-\frac{C+\Delta C}{n}+\frac{\phi\sqrt{q(1-q)}K}{\sqrt{n}}
\]

\red{With some algebra we get:}
\[
\left(\frac{B+C+(1+r)\Delta C}{\phi\sqrt{q(1-q)}K}\right)^2=n
\]
\end{proof}

\begin{remark}
\red{Value of \(n_{ZMPR}(\Delta C)\) (expression (\ref{eq:ZPn})) increases as }\(\Delta C\)\red{\ increases. We limit our analysis to positive capital change, so
value of \(n_{ZMPR}(\Delta C)\) is the smallest if }\(\Delta C=0\)\red{ and \(n_{ZMPR}(\Delta C)\) tends to infinity
as }\(\Delta C\)\red{\ tends to infinity. On the other hand the capital increase does not change the demand curve, and we know
that the demand curve is bounded. So if the point of intersection for zero capital
change (infinitesimally small) is under the demand curve there necessarily will be value of \(\Delta C\) for which the point of intersection will be on the demand curve (i.e. the three curve have a common point of intersection). It is very cumbersome
to investigate the uniqueness of this common point of intersection of the three curve. We cannot give a
precise mathematical proof of it, but an important fact that if we substitute expression (\ref{eq:ZPn}) and (\ref{eq:ZPP})
into the demand function we get a quadratic expression for \(\Delta C\). A quadratic expression can have 0, 1 or two roots.
If point \((P_{ZMPR}(0),n_{ZMPR}(0))\) is under the demand curve we know that at least on root has to be. Let we consider
\((P_{ZMPR}(\Delta C),n_{ZMPR}(\Delta C))\) as a curve. This curve starts from under the demand curve (\(\Delta C = 0\)) and
go above the demand curve. If it would go under the the demand curve for another \(\Delta C\) value, then it would have 
to go above again some further value of \(\Delta C\) (since \(n_{ZMPR}(\Delta C)\) tends to infinity), but it is not possible,
since it would mean three root of a quadratic expression. In pure mathematical analysis it could happen that the second (or first) point of intersection is only a point of tangency but we do not feel it realistic. In Section \ref{sec:exante} we will see, that
the three curve do not have a common point of intersection, or have exactly one. We checked numerically several
parameter set, and never find more than one root. We assume that there is at most one root exist. 
The premium in the point of intersection of the three curve is denoted by \(P_{1Z}\) (if it does not exists we consider it infinitely high)} 
\end{remark}

\red{Let us see the question of Nash equilibrium. The essential question whether condition \(P_{1Z} < \min(P_U,P_L)\) holds or not.
We discuss the two cases separately in Proposition \ref{prop:noadjcapNE} and Proposition \ref{prop:capadjNEc}.}

\begin{prop} \label{prop:noadjcapNE}
\red{If condition \(P_{1Z} \ge \min(P_U,P_L)\) holds than none of the companies will adjust its capital, we get back the situations
investigated in section \ref{sec:short}.}
\end{prop}

\begin{proof}
\red{The proof is quite trivial. To set premium lower than \(\min(P_U,P_L)\) need capital increase but with this increase, the firm
gets to a worse situation then in a Nash equilibrium described in Section \ref{sec:short}.}
\end{proof}

\begin{prop} \label{prop:capadjNEc}
\red{If condition \(P_{1Z} < \min(P_U,P_L)\) holds than pure Nash equilibrium does not exist if continuum-many
premium level are possible.}
\end{prop}

\begin{proof}
\red{Let us assume that the minimal premium is unique (only one company set this price). If it is higher than \(P_{1Z}\), then
any of the other companies will undercut. If it is \(P_{1Z}\) then it is worth to set a bit higher premium level. Let us assume
that at least two companies set the minimal price. Again, if it is higher than \(P_{1Z}\) one of the company will undercut. If it
would be \(P_{1Z}\), then it would cause loss for all of them, since getting a fraction of the market requires larger capital level
than the same fraction of a single company would require see Remark \ref{re:decreasing_ret_scale}.}
\end{proof}

\begin{prop} \label{prop:capadjNEd}
\red{If condition \(P_{1Z} < \min(P_U,P_L)\) holds than in a discrete setting one type of Nash equilibrium exist. One of the
companies set premium level \(P_{1Z}\), at least one of the others set the next available premium level.}
\end{prop}

\begin{proof}
\red{Let assume that one of the companies (let us call it first) set premium level \(P_{1Z}\), and at least on of the others set the next available premium level. It is easy to see that this is Nash equilibrium: if the first company sets higher premium level, two cases possibly: 1) It would increase it to the next premium level, but in this case it get only the half of the market
(or even less), if the possible prices are dense enough, it will cause it loss. 2) It would set an even higher premium level, then
it gets no market share. If some of other companies would set premium level \(P_{1Z}\), then it would cause loss for both of them
see Remark \ref{re:decreasing_ret_scale}.}
\end{proof}

\begin{remark}
\red{Two important remark about the Nash equilibrium described in Proposition \ref{prop:capadjNEd}:
\begin{itemize}
\item one of the companies gets the whole market
\item all of the companies suffer loss, if we consider the sunk cost of interest loss on initial capital level.
\end{itemize}
}
\end{remark}

\subsection{Ex ante decision on capital level} \label{sec:exante}

\red{In this section we also investigate decision on capital level, but in different content: we consider a two period game. In the first period the players
found companies and decide on capital level simultaneously. In the second period companies' capital level is fixed, and they decide on premium level.}

\red{To do this, we need some more preparation. First we have to describe how a monopoly behave in this setting.}

\begin{prop}
\red{In a monopolistic market the player found its company with capital }
\begin{equation} \label{eq:moncap}
C_{MC}=\frac{\left(\frac{\phi\sqrt{q(1-q)}K}{1+r}\right)^2-(\phi\sqrt{q(1-q)}K-\alpha)^2}{4qK}
\end{equation}
\red{and set premium level}
\begin{equation} \label{eq:moncapprem}
P_{MC}=\frac{2\alpha qK}{\alpha-(\phi\sqrt{q(1-q)}K)\frac{r}{1+r}}
\end{equation}
\red{at the second period.}
\end{prop}

\begin{proof}
\red{A monopoly wants to maximize its expected profit. In the second period it will set premium level }\(\max(P_M,P_U)\)\red{. It is quite clear
that }\(P_M\le P_U\) \red{ (otherwise the company would be founded with more capital then it would be necessary, the interest on the unnecessary part of
the capital decrease the company's potential profit). So the company never set lower premium than }\(P_M\)\red{ but it can be rational to set higher level:
setting higher level of premium in the second period means that the company can be founded with less capital, the gain in less interest loss can effect more in
profit than decrease caused by higher premium level than the monopolistic level. With algebraic calculation: since the company will set premium equals with }
\(P_U\)\red{, which is the intersection of MPR curve and demand function, the profit at this point will be:}
\begin{equation} \label{eq:moncapp}
\begin{array}{l}
\displaystyle
\frac{\left(-(\phi\sqrt{q(1-q)}K-\alpha)+\sqrt{(\phi\sqrt{q(1-q)}K-\alpha)^2+4qKC}\right)^2}{4q^2K^2} \times \\[2em]
\displaystyle
\hspace{2em}
\left(\frac{2\alpha q K}{-(\phi\sqrt{q(1-q)}K-\alpha)+\sqrt{(\phi\sqrt{q(1-q)}K-\alpha)^2+4qKC}} - qK\right) - 
rC
\end{array}
\end{equation}
\red{With some algebra we get (\ref{eq:moncapp}) equals with}
\begin{equation} \label{eq:moncapp2}
\frac{-2E^2-2\alpha E}{4qK} +
\frac{(2(\alpha+E)\sqrt{E^2+4qKC})-4(1+r)qKC}{4qK} \ ,
\end{equation}
\red{where }\(E=\phi\sqrt{q(1-q)}K-\alpha\)\red{. Derivative of (\ref{eq:moncapp2}) with respect to capital is:}
\begin{equation} \label{eq:moncapder}
\frac{\alpha+E}{\sqrt{E^2+4qKC}}-(1+r) \ .
\end{equation}
\red{First order condition gives }\(\frac{\left(\frac{E+\alpha}{r+1}\right)^2-E^2}{4qK}\)\red{. Substituting value of }\(E\)\red{ into the
first order condition we get (\ref{eq:moncap}). Substituting (\ref{eq:moncap}) into the demand function we get (\ref{eq:moncapprem}).}
\end{proof}

\red{Short analysis of expressions (\ref{eq:moncap}) and (\ref{eq:moncapprem}). Let we start with the case of 0 interest (}\(r=0\)\red{). In this case
expression (\ref{eq:moncapprem}) will equal }\(2qK\)\red{and the monopolistic player will found its company with capital level }
\[
=\frac{-\alpha^2+2\alpha\phi \sqrt{q(1-q)}K}{4qK} \ ,
\]
\red{which is just as big, as premium level }\(P_M=2qK\)\red{ requires. As the interest rate start increase monopolistic capital start to increase as well, which coincide with a premium increase. Increasing
interest rate we will reach a value}\(r=\frac{\alpha}{\phi\sqrt{q(1-q)}K-\alpha}\)\red{ when denominator of (\ref{eq:moncapprem}) become 0. It means
if }\(r\ge\frac{\alpha}{\phi\sqrt{q(1-q)}K-\alpha}\)\red{ then the interest rate is so high, that no insurance company can operate in the market without financial
loss.}

\red{A monopolistic player founds its company
with capital level \(C_{MC}\). Let us consider the MPR curve which belongs to capital level \(C_{MC}\). This MPR curve
intersects the half demand at premium level \(P_{MCL}\)}

\red{For determining Nash equilibrium points we need to revise the zero-profit curve. Zero profit curves was introduced in Section \ref{sec:mod}. The concept remain
unchanged but we have to adjust it to current situation. In our context zero profit curve is:}
\[
ZP(P, C)=\frac{r C}{P-qK}
\] 
 
\red{Again, the zero profit curve and the MPR curve has a unique point of intersection (for fixed \(C\)) where the number
of policies is}
\begin{equation} \label{eq:ZPn2}
n_{ZMPR}(C)\left(\frac{(1+r) C}{\phi\sqrt{q(1-q)}K}\right)^2
\end{equation}
\red{end the premium level is}
\begin{equation} \label{eq:ZPP2}
P_{ZMPR}(C)=qK+(\phi\sqrt{q(1-q)}K)^2\frac{r}{(1+r)^2} \frac{1}{C} \ .
\end{equation}

\red{In this case if we substitute  \(P_{ZMPR}(C)\) and \(n_{ZMPR}(C)\) into the demand function we get a unique value
for \(C\). We denote this level of capital with \(C_{1Z}\), and \(P_{1ZU}=P_{ZMPR}(C_{1Z})\). Considering the MPR curve,
which go through point (\(P_{ZMPR}(C_{1Z}),n_{ZMPR}(C_{1Z})\)). This MPR curve intersects the half demand curve at premium level
\(P_{1ZL}\). Any point in the demand curve \green{right} to \(P_{1ZU}\) will result positive profit, and any point in the demand curve
\green{left} to \(P_{1ZU}\) will result negative profit.}

\red{Similarily: }\(C_{2Z}\)\red{denote the capital level at which the half inverse demand curve, the zero profit curve and }\(MPR\)\red{ curve
intersects at the same point }(\(P_{2ZL}\))\red{. This level of capital is also unique. We can state that
\(P_{1ZU} < P_{2ZL}\) regardless
whether MPR curve intersects the inverse demand curve at its decreasing or increasing part.}

\red{We investigate a two period model. In first period the companies set its capital level, in the second period they set premium
level based on fixed level of capital. We analyzed in section \ref{sec:short} the possible Nash equilibrium assuming fixed level of capital.
In certain cases we get multiple Nash equilibrium, in other cases there was not Pure equilibrium or we get different results
for continuous and discrete settings. We have to decide what payoff will get the companies is the second period assuming
fixed level of capital. We assume that companies considering the worst case scenario in the second period. The worst case
scenario means the Nash equilibrium with the lowest premium level (if the premium level is above the monopolistic price,
not necessarily the lowest premium is the worst case but we feel this situation not relevant in our case).}

\begin{prop}
\red{If \(P_{MC} < P_{2ZL}\) and \(P_{MCL}< P_{1ZU}\) then a pure Nash equilibrium exist: one of the companies found its company with
capital level \(C_{MC}\), the other with capital level 0 (i.e. do not found its company). In the second period
the monopolistic company set premium \(P_{MC}\).}
\end{prop}

\begin{proof}
\red{It is quite clear, that the monopolistic company does not want to differ from its strategy. We have to see that it is
not optimal for the other company to enter the market. It is quite clear, that the other player will found its company, if it gets some market share.
To get the whole market in the second period it has to set premium level lower than \(P_{MCL}\) in the second period. To do
this it requires more capital than \(C_{1Z}\)  (since \(P_{MCL}< P_{1ZU}\)), but such large level of capital at a premium level lower than \(P_{1ZU}\)
would cause loss to the company. With lower level of capital it is still possible to get the half of the market, but such
level of capital still has to be higher than \(C_{2Z}\) (since \(P_{MC} < P_{2ZL}\)) which also cause loss to the company.}

\end{proof}

\begin{prop}\label{prop:nane1}
\red{If \(P_{MC} \ge P_{2ZL}\) then there is no pure Nash equilibrium.}
\end{prop}

\begin{proof}
\green{According to  Remark \ref{re:decreasing_ret_scale}. a market that is twice as large can be covered with less than twice as much capital, so if the half of the market results 0 profit, it is worthwhile to serve the whole market at a lower price even with a capital increase, so a positive profit can be achieved. The cost of raising capital is less than the increase in income. Three different cases can occur:} 
\begin{itemize}

	\item \green{The companies have the same level of market share (with same or different capital level it is also possible).
	The worst case premium in the second period means premium level \(P_{2*} \ge P_{2ZL}\). It is worth \red{(at least) to the larger company} to increase its level of capital to a point, where it gets the whole market. Insuring twice as many clients at premium level \red{infinitesimally less than} \(P_{2ZL}\) requires
	less than twice as large capital level, so this new strategy result higher profit for the company.}

	\item \green{The players found their companies with different positive level of capital and the bigger company gets the whole market. In this case the smaller company should decrease its capital level to zero, so it can reach higher profit.}
	
	\item \green{The players found their companies with different level of capital, the bigger company gets the whole market and the smaller company's capital is zero. If only one company exist in the market its capital level has to be \(C_{MC}\). But in this case the smaller company gets at least the half of the market, if it enters to market with capital level \(C_{Z2}\), and this strategy do not cause loss for it.}
\end{itemize}
\end{proof}

\begin{prop}\label{prop:nona2}
\red{If \(P_{MCL} \ge P_{1Z\green{U}}\) then there is no pure Nash equilibrium.}
\end{prop}

\begin{proof}
\red{A monopolistic company would be founded at capital level \(C_{MC}\) \green{with price \(P_{MC}\)}. But it is worth to undercut in this case \green{in iterval \([P_{1ZU}, P_{MC}]\), where the profit is positive.} If two companies
are in the market \green{and} its capital level equal then it has to be less than \green{\(C_{2Z}\)}; one of the company will undercut. \green{With a higher level of capital it can offer a lower price, which is profitable according to Remark \ref{re:decreasing_ret_scale}.} If its
capital level differ, two cases are possible. 1) Both the companies
get the half of the market. In this case the profit is less for the bigger company; it could increase its profit with decreasing the capital level. 2)The bigger company gets the whole market. In this case the smaller company would decrease its capital level to zero.}
\end{proof}

\red{In previous section we found, that in a discrete settings we found Nash equilibrium. It is not hard to see, that proof of
Proposition \ref{prop:nane1} and \ref{prop:nona2} also work in a discrete setting if the possible capital levels are dense enough.}

\begin{table}[ht]
	\caption{Parameters of the simulations} 
	\centering 
	\begin{tabular}{c|c|c|c|c } 
		\hline\hline 
		Parameters & Original & Minimum & Maximum &Step  \\ [0.5ex] 
		\hline 
		
		$\alpha$ & 150 & 90 & 200 &20\\ 
		$q$ & 0.025 & 0.01&0.2&0.04 \\
		$K$ & 600&100&1000&200 \\
		$r$ & 0.2&0.01&0.3&0.05 \\
		  [1ex] 
		\hline 
	\end{tabular}
	\label{table:par_values} 
\end{table}

\green{Analysing mixed strategies is a quite complex problem in this case. To get an impression of existing Nash equilibrium strategies
we investigated the question as a bimatrix game. Both player can found its company with capital level \(C_1,C_2,\dots,C_n\), \red{where \(C_1=0, C_n=C_{1Z}\), and other capital levels are distributed uniformly between these two values}}. \green{Mixed Nash equilibrium strategies calculated with method bimatrix-solver-master\footnote{https://cgi.csc.liv.ac.uk/~rahul/bimatrix\_solver/}. Unfortunately the method works with limited payoff matrix. We could solve the problem within one hour for at most 20x20 matrix size.} \red{We listed all parameter tuples listed in table \ref{table:par_values} and considered only the cases where MPR curves intersects inverse demand curves for all investigated capital levels in its decreasing or increasing part. There are 540 cases where all MPR curves intersect the inverse demand curve in its increasing part and none when all MPR curves intersects the inverse demand curve in its decreasing part. We calculated all Nash equilibrium. We found three types of Nash equilibria:} 

\begin{enumerate}

\item \red{Both player found its company, \green{the probability of zero capital level is zero}; the expected payoff is positive for both players. Insurance market surely exist in this case.}

\item \red{One of the player founds surely its company (with positive capital level), and its expected payoff is positive.
The other player may not found its company (plays with positive probability the strategy belongs to zero capital level) and its expected payoff is zero. The monopolistic price also appear with positive probability.}

\item \red{Both player may not found its company (plays with positive probability the 0 capital level strategy), and the expected payoff is zero for both player. It also means that there is no insurance market with positive probabilities. On the other hand the monopolistic premium (or even higher) will
appear with positive probability.}
\end{enumerate}

Distribution of Nash equilibria amongst the three types are: Type 1: 326, type 2: 976, type 3: 564.

\red{Derivation of payoff matrices and and a numerical example for the three possible equilibrium types described in Appendix D}

\section{Conclusion} \label{sec:conc}

The specific objective of this paper was to study the effect of the Solvency II directive's capital requirement in the market equilibrium in a Bertrand oligopoly market. In the traditional Bertrand game the profit of the companies is zero, even if there are only two companies in the market. This research has shown that introducing the solvency capital requirement constraint \green{can lead to different market anomalies}. Assuming homogenous fixed capital level a continuum of symmetric Nash equilibria occur. The equilibrium premiums are not less than the net premium. At net premium the expected profit is the fixed cost, if the premium is higher, then the technical result becomes positive. In some cases a subinterval of the Nash equilibrium interval can ensure positive expected profit for the insurance companies. If the interest rate is lower,  positive profit occurs more often. 

There are three notable premium levels: the two endpoints of the possible equilibrium interval and  $2qK$ which gives the highest expected profit for the companies if it satisfies the solvency capital requirement. The lowest possible equilibrium premium is not less than the net premium. The consumers can buy insurance at the cheapest premium at that level, while the demand is the highest in this case. The highest possible equilibrium premium is also a special premium level, at this premium even one company can serve the whole market, thus there is no risk to pay penalty because of the solvency capital requirement. If  $2qK$ is in the equilibrium interval, this premium level is the best for the insurance companies. If the $2qK$ is lower than all the possible equilibrium prices than the lowest possible premium gives the highest expected profit. If it is higher than all the possible equilibrium prices then the highest endpoint of the interval gives the largest expected profit.    

The increasing of the number of companies causes decreasing in the lower endpoint of the equilibrium interval, while the higher endpoint is unchanged. If the level of capital is increasing both of the endpoints are decreasing, if they are above the net premium. If the total capital in the market is fixed, the increasing of the number of companies leads to a higher lower endpoint of the equilibrium interval. Usually allowing mergers in the market causes higher equilibrium price, but there are some parameter combinations, when the monopoly market with the total capital level sells contracts at a lower premium. One company with a higher level of capital can ensure lower possible equilibrium premiums. This is a similar case to the natural monopoly market. Higher level of capital can ensure lower premium level, but holding capital has a fixed cost, so one company with higher capital can sell at a lower possible equilibrium premium.

\green{Assuming heterogen capital level there are different types of equlibria. If the decreasing part of the MPR curve is relevant, the larger company is on the market alone. In same cases the equlibrium is not unique. Introducung the capital decision the possibility of positive profit, equilibrium with one company and market failure is positive.}

The scope of this study was limited in terms of the simplifying assumptions of the model. Greater efforts are needed to derive these results in a more general setting and to show the conditions under which the presented phenomena occur. Lifting the main assumptions about the short run model -- where the level of capital is not a decision variable --, the homogeneous companies and the simultaneous decision making can also lead to important results. A further research topic can be the case, when the insurers have beliefs about the behavior of each other, and they also take this into account in their decisions.

\newpage

\section*{Appendix A}
A higher level of MPR than $K$ is meaningless. \red{This happens if}
$$  qK-\frac{C}{n}+\frac{\phi\sqrt{q(1-q)}K}{\sqrt{n}}>K \ . $$
\red{The condition is satisfied if:}
$$K(n(q-1)+\phi\sqrt{n}\sqrt{q(1-q)})-C)>0 \ ,$$
\red{which necessarily will not be true if $n$ is large enough.}

\section*{Appendix B}
The proof of Proposition \ref{prop:intersection}. The intersection of the inverse demand function and the MPR curve:
\[
\frac{\alpha}{\sqrt{n}} = qK-\frac{C}{n}+\frac{\phi\sqrt{q(1-q)}K}{\sqrt{n}}
\]

After rearranging the term, we get:
\[
0=qKn+\sqrt{n}(\phi\sqrt{q(1-q)}K-\alpha)-C \ ,
\]
which is a quadratic expression in \(\sqrt{n}\). Using the quadratic formula for roots:
\[
(\sqrt{n})_{1,2}=\frac{-(\phi\sqrt{q(1-q)}K-\alpha)\pm\sqrt{(\phi\sqrt{q(1-q)}K-\alpha)^2+4qKC}}{2qK}
\]

The discriminant is always positive, so there are two roots. However, from expression
\[
(\sqrt{n})_{1,2}=\frac{-(\phi\sqrt{q(1-q)}K-\alpha)-\sqrt{(\phi\sqrt{q(1-q)}K-\alpha)^2+4qKC}}{2qK}
\]
we get a negative value for \(\sqrt{n}\), which is out of context. So the intersection point of the two function
is at
\[
n=\frac{\left(-(\phi\sqrt{q(1-q)}K-\alpha)+\sqrt{(\phi\sqrt{q(1-q)}K-\alpha)^2+4qKC}\right)^2}{4q^2K^2}
\]

The premium at the intersection point is
\[
P=\frac{2\alpha qK}{-(\phi\sqrt{q(1-q)}K-\alpha)+\sqrt{(\phi\sqrt{q(1-q)}K-\alpha)^2+4qKC}}
\]

\section*{Appendix C}

Complete proof of Proposition \ref{prop:incr_comp_fixC}:

Multiplying both the numerator and denominator of (\ref{eq:P_L}) by \(\sqrt{I}\) we get:
\begin{equation} \label{eq:AB}
P_L^C(I)=\frac{2\alpha qK}{-(\sqrt{I}\phi\sqrt{q(1-q)}K-\alpha)+\sqrt{(\sqrt{I}\phi\sqrt{q(1-q)}K-\alpha)^2+4qKC}} \ ,
\end{equation}
 
The numerator of (\ref{eq:AB}) does not depend on \(I\), it is enough to consider the denominator. Denote it by \(D(I)\). The derivative of \(D(I)\) is
\[
\begin{array}{l}
\displaystyle
\frac{d D(I)}{d I}=\\
\displaystyle
\hspace{2em}-0.5\frac{\phi\sqrt{q(1-q)}K}{\sqrt{I}}+\\
\displaystyle
\hspace{3em}0.5\frac{2(\sqrt{I}\phi\sqrt{q(1-q)}K-\alpha)}{\sqrt{(\sqrt{I}\phi\sqrt{q(1-q)}K-\alpha)^2+4qKC}}0.5\frac{\phi\sqrt{q(1-q)}K}{\sqrt{I}}=\\
\displaystyle
0.5\frac{\phi\sqrt{q(1-q)}K}{\sqrt{I}}\left(\frac{(\sqrt{I}\phi\sqrt{q(1-q)}K-\alpha)}{\sqrt{(\sqrt{I}\phi\sqrt{q(1-q)}K-\alpha)^2+4qKC}}-1\right) \ .
\end{array}
\]
 
It is easy to see that
\[
\frac{(\sqrt{I}\phi\sqrt{q(1-q)}K-\alpha)}{\sqrt{(\sqrt{I}\phi\sqrt{q(1-q)}K-\alpha)^2+4qKC}} < 1 \ ,
\]
implying that \(\frac{d D(I)}{d I}<0\), so \(D(I)\) is a decreasing function of \(I\). If \(D(I)\) is a decreasing function
then \(P_L^C(I)\) is an increasing function of \(I\); as the number of insurance companies increases, so does \(P_L^C\).

\section*{Appendix D}

\red{{\it Derivation of payoff matrices.} Since during our simulation we do not find any parameter settings where all of the MPR curves intersect the inverse demand curve on its decreasing part, so we investigate only the cases where all of the MPR curves intersect the inverse demand curve in its increasing part. If both company found its company with the same level of capital (which is not lower than \green{\(C_{1Z}\))}, lowest premium level in the second period is \(P_L\) described in Proposition \ref{prop:NE}. If players found its company with different level of capital, we have to distinguish two possibilities: if \(_SP_L \le {_HP_U}\) then the lowest premium level in the second period is \(_SP_L\) (see Proposition \ref{prop:ACLIIc}) and both of the companies get half of the market. If \(_HP_U < {_SP_L}\) than the larger company gets the whole market at price \(_HP_U\) (see Proposition \ref{prop:ACLIIcdD}). If a player do not found company \green{($C=0$)}, its payoff is 0. If there is only one company in the market is the second period then it will set monopolistic price except its capital level is insufficient for it. So the payoff is \(\max(P_M,P_U)\)}.

In table \ref{table:payoff} and table \ref{table:payoff2} we give two payoff matrices (of the row player). The first belong to parameter settings $q=0.05, K=900, \alpha=90, r=0.01$, the second to paremeters $q=0.025, K=600, \alpha=150, r=0.2$.

In table \ref{table:eq} we give Nash equilibrium strategies for payoff matrix appear in table \ref{table:payoff}. We can see, that in all the three cases the expected payoff is positive for both players. \green{It is an example for equilibrium type 1, when both companies are in the market and the expected payoff is positive.}

In table \ref{table:eq2} we give Nash equilibrium strategies for payoff matrix appear in table \ref{table:payoff2}. \green{In this case two type of Nash equilibrium exist. Equilibrium 4,5 and 6 is an example for equilibrium type 3, both players expected payoff is 0 and both of them will be away from the market with positive probability (so the probability of market failure is also positive). Equilibrium 1,2,3,7,8 and 9 are the 2nd type of Nash equilibrium, where one of the players gets positive expected payoff.}

\newgeometry{left=3cm,bottom=0.1cm, top=0.1cm}
\begin{sidewaystable}[ht]
	\caption{Payoff matrix} 
	\centering 
	\tiny
	\begin{tabular}{c|c|c|c|c|c|c|c|c|c|c|c|c|c|c|c|c|c|c|c|c|c|c } 
		\hline\hline 
		Capital level&52.5&102&151.5&201&250.5&300&349.5&399&448.5&498&547.5&597&646.5&696&745.5&795&844.5&894&943.5&0\\ [0.5ex] 
		\hline 

52.5&6.33&-0.52&-0.52&-0.52&-0.52&-0.52&-0.52&-0.52&-0.52&-0.52&-0.52&-0.52&-0.52&-0.52&-0.52&-0.52&-0.52&-0.52&-0.52&10\\
102&17.96&11.06&-1.02&-1.02&-1.02&-1.02&-1.02&-1.02&-1.02&-1.02&-1.02&-1.02&-1.02&-1.02&-1.02&-1.02&-1.02&-1.02&-1.02&17.96\\
151.5&24.58&24.58&14.66&14.66&-1.52&-1.52&-1.52&-1.52&-1.52&-1.52&-1.52&-1.52&-1.52&-1.52&-1.52&-1.52&-1.52&-1.52&-1.52&24.58\\
201&29.94&29.94&14.16&17.19&17.19&-2.01&-2.01&-2.01&-2.01&-2.01&-2.01&-2.01&-2.01&-2.01&-2.01&-2.01&-2.01&-2.01&-2.01&29.94\\
250.5&34.13&34.13&34.13&16.69&18.71&18.71&-2.51&-2.51&-2.51&-2.51&-2.51&-2.51&-2.51&-2.51&-2.51&-2.51&-2.51&-2.51&-2.51&34.13\\
300&37.22&37.22&37.22&37.22&18.22&19.29&19.29&19.29&-3&-3&-3&-3&-3&-3&-3&-3&-3&-3&-3&37.22\\
349.5&39.27&39.27&39.27&39.27&39.27&18.79&18.96&18.96&18.96&-3.5&-3.5&-3.5&-3.5&-3.5&-3.5&-3.5&-3.5&-3.5&-3.5&39.27\\
399&40.34&40.34&40.34&40.34&40.34&18.29&18.46&17.78&17.78&17.78&17.78&-4&-4&-4&-4&-4&-4&-4&-4&40.34\\
448.5&40.49&40.49&40.49&40.49&40.49&40.49&17.97&17.29&15.79&15.79&15.79&15.79&-4.49&-4.49&-4.49&-4.49&-4.49&-4.49&-4.49&40.49\\
498&39.76&39.76&39.76&39.76&39.76&39.76&39.76&16.79&15.3&13.03&13.03&13.03&13.03&-4.99&-4.99&-4.99&-4.99&-4.99&-4.99&40.01\\
547.5&38.19&38.19&38.19&38.19&38.19&38.19&38.19&16.29&14.8&12.54&9.54&9.54&9.54&9.54&9.54&-5.48&-5.48&-5.48&-5.48&39.52\\
597&35.84&35.84&35.84&35.84&35.84&35.84&35.84&35.84&14.31&12.04&9.04&5.34&5.34&5.34&5.34&5.34&-5.98&-5.98&-5.98&39.02\\
646.5&32.73&32.73&32.73&32.73&32.73&32.73&32.73&32.73&32.73&11.54&8.54&4.84&0.47&0.47&0.47&0.47&0.47&0.47&-6.47&38.53\\
696&28.9&28.9&28.9&28.9&28.9&28.9&28.9&28.9&28.9&28.9&8.05&4.35&-0.03&-5.05&-5.05&-5.05&-5.05&-5.05&-5.05&38.03\\
745.5&24.38&24.38&24.38&24.38&24.38&24.38&24.38&24.38&24.38&24.38&7.55&3.85&-0.53&-5.54&-11.18&-11.18&-11.18&-11.18&-11.18&37.53\\
795&19.2&19.2&19.2&19.2&19.2&19.2&19.2&19.2&19.2&19.2&19.2&3.35&-1.02&-6.04&-11.67&-17.9&-17.9&-17.9&-17.9&37.04\\
844.5&13.4&13.4&13.4&13.4&13.4&13.4&13.4&13.4&13.4&13.4&13.4&13.4&-1.52&-6.53&-12.17&-18.39&-25.19&-25.19&-25.19&36.54\\
894&6.99&6.99&6.99&6.99&6.99&6.99&6.99&6.99&6.99&6.99&6.99&6.99&-2.01&-7.03&-12.66&-18.89&-25.68&-33.02&-33.02&36.05\\
943.5&0&0&0&0&0&0&0&0&0&0&0&0&0&-7.53&-13.16&-19.38&-26.18&-33.52&-41.38&35.55\\
0&0&0&0&0&0&0&0&0&0&0&0&0&0&0&0&0&0&0&0&0\\

		\hline 
	\end{tabular}
	\label{table:payoff} 
\end{sidewaystable}
\restoregeometry

\newgeometry{left=3cm,bottom=0.1cm, top=0.1cm}
\begin{sidewaystable}[ht]
	\caption{Payoff matrix} 
	\centering 
	\tiny
	\begin{tabular}{c|c|c|c|c|c|c|c|c|c|c|c|c|c|c|c|c|c|c|c|c|c|c } 
		\hline\hline 
		Capital level&58.87&137.36&215.85&294.33&372.82&451.31&529.8&608.29&686.78&765.27&843.76&922.25&1000.74&1079.23&1157.71&1236.2&1314.69&1393.18&1471.67&0\\ [0.5ex] 
		\hline 

	58.87&37.45&-15.49&-15.49&-15.49&-15.49&-15.49&-15.49&-15.49&-15.49&-15.49&-15.49&-15.49&-15.49&-15.49&-15.49&-15.49&-15.49&-15.49&-15.49&89.19\\
	137.36&147.75&62.22&-30.98&-30.98&-30.98&-30.98&-30.98&-30.98&-30.98&-30.98&-30.98&-30.98&-30.98&-30.98&-30.98&-30.98&-30.98&-30.98&-30.98&147.75\\
	215.85&187.34&187.34&77.43&-46.47&-46.47&-46.47&-46.47&-46.47&-46.47&-46.47&-46.47&-46.47&-46.47&-46.47&-46.47&-46.47&-46.47&-46.47&-46.47&187.34\\
	294.33&213.69&213.69&213.69&85.09&-61.97&-61.97&-61.97&-61.97&-61.97&-61.97&-61.97&-61.97&-61.97&-61.97&-61.97&-61.97&-61.97&-61.97&-61.97&213.69\\
	372.82&230.11&230.11&230.11&230.11&86.58&-77.46&-77.46&-77.46&-77.46&-77.46&-77.46&-77.46&-77.46&-77.46&-77.46&-77.46&-77.46&-77.46&-77.46&230.11\\
	451.31&238.74&238.74&238.74&238.74&238.74&82.92&82.92&-92.95&-92.95&-92.95&-92.95&-92.95&-92.95&-92.95&-92.95&-92.95&-92.95&-92.95&-92.95&238.74\\
	529.8&241.03&241.03&241.03&241.03&241.03&67.43&74.87&74.87&-108.44&-108.44&-108.44&-108.44&-108.44&-108.44&-108.44&-108.44&-108.44&-108.44&-108.44&241.03\\
	608.29&238.05&238.05&238.05&238.05&238.05&238.05&59.38&63.01&63.01&-123.93&-123.93&-123.93&-123.93&-123.93&-123.93&-123.93&-123.93&-123.93&-123.93&238.05\\
	686.78&230.58&230.58&230.58&230.58&230.58&230.58&230.58&47.52&47.83&47.83&47.83&-139.42&-139.42&-139.42&-139.42&-139.42&-139.42&-139.42&-139.42&230.58\\
	765.27&219.23&219.23&219.23&219.23&219.23&219.23&219.23&219.23&32.34&29.69&29.69&29.69&-154.91&-154.91&-154.91&-154.91&-154.91&-154.91&-154.91&219.23\\
	843.76&204.49&204.49&204.49&204.49&204.49&204.49&204.49&204.49&16.84&14.2&8.91&8.91&8.91&8.91&-170.4&-170.4&-170.4&-170.4&-170.4&204.6\\
	922.25&186.75&186.75&186.75&186.75&186.75&186.75&186.75&186.75&186.75&-1.29&-6.58&-14.25&-14.25&-14.25&-14.25&-185.9&-185.9&-185.9&-185.9&189.1\\
	1000.74&166.33&166.33&166.33&166.33&166.33&166.33&166.33&166.33&166.33&166.33&-22.07&-29.74&-39.56&-39.56&-39.56&-39.56&-39.56&-201.39&-201.39&173.61\\
	1079.23&143.49&143.49&143.49&143.49&143.49&143.49&143.49&143.49&143.49&143.49&-37.56&-45.23&-55.05&-66.84&-66.84&-66.84&-66.84&-66.84&-216.88&158.12\\
	1157.71&118.47&118.47&118.47&118.47&118.47&118.47&118.47&118.47&118.47&118.47&118.47&-60.72&-70.54&-82.33&-95.93&-95.93&-95.93&-95.93&-95.93&142.63\\
	1236.2&91.46&91.46&91.46&91.46&91.46&91.46&91.46&91.46&91.46&91.46&91.46&91.46&-86.03&-97.82&-111.42&-126.69&-126.69&-126.69&-126.69&127.14\\
	1314.69&62.61&62.61&62.61&62.61&62.61&62.61&62.61&62.61&62.61&62.61&62.61&62.61&-101.52&-113.32&-126.91&-142.18&-158.98&-158.98&-158.98&111.65\\
	1393.18&32.09&32.09&32.09&32.09&32.09&32.09&32.09&32.09&32.09&32.09&32.09&32.09&32.09&-128.81&-142.41&-157.67&-174.47&-192.71&-192.71&96.16\\
	1471.67&0&0&0&0&0&0&0&0&0&0&0&0&0&0&-157.9&-173.16&-189.96&-208.2&-227.77&80.67\\
	0&0&0&0&0&0&0&0&0&0&0&0&0&0&0&0&0&0&0&0&0\\

		\hline 
	\end{tabular}
	\label{table:payoff2} 
\end{sidewaystable}
\restoregeometry

\newgeometry{left=3cm,bottom=0.1cm, top=0.1cm}
\begin{sidewaystable}[ht]
	\caption{Equilibria} 
	\centering 
	\tiny
	\begin{tabular}{c|c|c|c|c|c|c|c|c|c|c|c|c|c|c|c|c|c|c|c|c|c|c|c|c } 
		\hline\hline 
		Equilibrium&Players/Strategies&52.5&102&151.5&201&250.5&300&349.5&399&448.5&498&547.5&597&646.5&696&745.5&795&844.5&894&943.5&0&Expected payoff\\ [0.5ex] 
		\hline 

	1&2&0&0&0&0&0&0&0&0.099&0.096&0.105&0.297&0.161&0.098&0.144&0&0&0&0&0&0&8.4\\
	1&1&0&0&0&0&0&0&0&0&0.11&0.125&0.334&0.066&0.094&0.132&0&0.139&0&0&0&0&11.026\\
	2&2&0&0&0&0&0&0&0&0.025&0.106&0.118&0.355&0.067&0.095&0.136&0&0.098&0&0&0&0&9.149\\
	2&1&0&0&0&0&0&0&0&0.025&0.106&0.118&0.355&0.067&0.095&0.136&0&0.098&0&0&0&0&9.149\\
	3&2&0&0&0&0&0&0&0&0&0.11&0.125&0.334&0.066&0.094&0.132&0&0.139&0&0&0&0&11.026\\
	3&1&0&0&0&0&0&0&0&0.099&0.096&0.105&0.297&0.161&0.098&0.144&0&0&0&0&0&0&8.4\\

		\hline 
	\end{tabular}
	\label{table:eq} 
\end{sidewaystable}
\restoregeometry


\newgeometry{left=3cm,bottom=0.1cm, top=0.1cm}
\begin{sidewaystable}[ht]
	\caption{Equilibria} 
	\centering 
	\tiny
	\begin{tabular}{c|c|c|c|c|c|c|c|c|c|c|c|c|c|c|c|c|c|c|c|c|c|c|c|c } 
		\hline\hline 
		Equilibrium&Players/Strategies&58.87&137.36&215.85&294.33&372.82&451.31&529.8&608.29&686.78&765.27&843.76&922.25&1000.74&1079.23&1157.71&1236.2&1314.69&1393.18&1471.67&0&Expected payoff\\ [0.5ex] 
		\hline 

	1&2&0&0&0&0&0&0&0&0&0&0&0.21&0.12&0.27&0&0.11&0.12&0&0&0&0.18&0\\
	1&1&0.17&0&0&0&0&0&0&0&0&0&0.23&0.13&0.24&0&0.11&0.12&0&0&0&0&3\\
	2&2&0&0.01&0&0&0&0&0&0&0&0&0.22&0.12&0.27&0&0.11&0.12&0&0&0&0.15&0\\
	2&1&0.17&0&0&0&0&0&0&0&0&0&0.23&0.13&0.23&0&0.11&0.12&0&0.01&0&0&0.2\\
	3&2&0&0&0&0&0&0&0&0&0&0&0.19&0.12&0.09&0&0.1&0.11&0&0.02&0&0.37&0\\
	3&1&0&0&0.12&0&0&0&0&0&0&0.05&0.22&0.13&0.31&0&0.11&0.07&0&0&0&0&38.94\\
	4&2&0.03&0&0&0&0&0&0&0&0&0&0.22&0.12&0.27&0&0.11&0.12&0&0&0&0.13&0\\
	4&1&0.03&0&0&0&0&0&0&0&0&0&0.22&0.12&0.27&0&0.11&0.12&0&0&0&0.13&0\\
	5&2&0&0.01&0&0&0&0&0&0&0&0&0.22&0.12&0.27&0&0.11&0.12&0&0&0&0.15&0\\
	5&1&0.05&0&0&0&0&0&0&0&0&0&0.21&0.12&0.26&0&0.11&0.12&0&0.01&0&0.12&0\\
	6&2&0.05&0&0&0&0&0&0&0&0&0&0.21&0.12&0.26&0&0.11&0.12&0&0.01&0&0.12&0\\
	6&1&0&0.01&0&0&0&0&0&0&0&0&0.22&0.12&0.27&0&0.11&0.12&0&0&0&0.15&0\\
	7&2&0.17&0&0&0&0&0&0&0&0&0&0.23&0.13&0.23&0&0.11&0.12&0&0.01&0&0&0.2\\
	7&1&0&0.01&0&0&0&0&0&0&0&0&0.22&0.12&0.27&0&0.11&0.12&0&0&0&0.15&0\\
	8&2&0.17&0&0&0&0&0&0&0&0&0&0.23&0.13&0.24&0&0.11&0.12&0&0&0&0&3\\
	8&1&0&0&0&0&0&0&0&0&0&0&0.21&0.12&0.27&0&0.11&0.12&0&0&0&0.18&0\\
	9&2&0&0&0.12&0&0&0&0&0&0&0.05&0.22&0.13&0.31&0&0.11&0.07&0&0&0&0&38.94\\
	9&1&0&0&0&0&0&0&0&0&0&0&0.19&0.12&0.09&0&0.1&0.11&0&0.02&0&0.37&0\\

		\hline 
	\end{tabular}
	\label{table:eq2} 
\end{sidewaystable}
\restoregeometry

\end{document}